\documentclass[a4paper]{article}
\usepackage[margin=0.75in]{geometry}

\usepackage{tocloft}
\usepackage{graphicx}
\usepackage{amsmath}
\usepackage{amsfonts}
\usepackage{mathrsfs}
\usepackage[all,cmtip]{xy}
\usepackage{soul}
\usepackage{epigraph}
\usepackage{hyperref}
\usepackage{amsthm}

\newcounter{theorem}
\newtheorem{thm}{Theorem}[section]

\newtheorem{lemma}[theorem]{Lemma}

\title{The Poincar\'{e}-Boltzmann Machine: from Statistical Physics to Machine Learning and back}
\author{Pierre Baudot \\  Median Technologies, Sophia Antipolis and Inserm UNIS UMR1072 - Université Aix-Marseille\\
	pierre.baudot@gmail.com}

\date{5th July 2019}

\begin{document}
	\maketitle
	
	\begin{abstract}
		This paper presents the computational methods of information cohomology applied to genetic expression in \cite{Tapia2018,Baudot2019,Baudot2018} and in the companion paper \cite{Baudot2019} and proposes its interpretations in terms of statistical physics and machine learning. In order to further underline the Hochschild cohomological nature af information functions and chain rules, following \cite{Baudot2015a,Baudot2015,Vigneaux2017}, the computation of the cohomology in low degrees is detailed to show more directly that the $k$ multivariate mutual-informations ($I_k$) are $k$-coboundaries. The $k$-cocycles condition corresponds to $I_k=0$, which generalize statistical independence to arbitrary dimension $k$ \cite{Baudot2019}. Hence the cohomology can be interpreted as quantifying the statistical dependences and the obstruction to factorization. The topological approach allows to investigate Shannon's information in the multivariate case without the assumptions of independent identically distributed variables and of statistical interactions without mean field approximations. We develop the computationally tractable subcase of simplicial information cohomology represented by entropy $H_k$ and information $I_k$ landscapes and their respective paths. The $I_1$ component defines a self-internal energy functional $U_k$, and $(-1)^k I_{k,k\geq2}$ components define the contribution to a free energy functional $G_k$ (the total correlation) of the k-body interactions. The set of information paths in simplicial structures is in bijection with the symmetric group and random processes, provides a trivial topological expression of the 2nd law of thermodynamic. The local minima of free-energy, related to conditional information negativity, and conditional independence, characterize a minimum free energy complex. This complex formalizes the minimum free-energy principle in topology, provides a definition of a complex system, and characterizes a multiplicity of local minima that quantifies the diversity observed in biology. I give an interpretation of this complex in terms of frustration in glass and of  Van Der Walls k-body interactions for data points.
	\end{abstract}

\epigraph{\textit{"Now what is science? ...it is before all a classification, a manner of bringing together facts which appearances separate, though they are bound together by some natural and hidden kinship. Science, in other words, is a system of relations. ...it is in relations alone that objectivity must be sought. ...it is relations alone which can be regarded as objective. External objects... are really objects and not fleeting and fugitive appearances, because they are not only groups of sensations, but groups cemented by a constant bond. It is this bond, and this bond alone, which is the object in itself, and this bond is a relation."}}{H. Poincar\'{e}}

\tableofcontents

\section{Introduction} 

The present paper has no deep mathematical or novelty pretension and just aims to present some computational aspects and applications to data of the simplicial simplest special case of the cohomology theory developed in  \cite{Baudot2015a,Baudot2019} and extended by Vigneaux \cite{Vigneaux2017, Vigneaux2019} to Tsallis entropies and differential entropy (etc.). The direct application to genetic expression measures, performance in cell type or gene module detection and its relevance in epigenetic co-regulation and differentiation terms can be find in \cite{Tapia2018}. In its application to empirically measured data, information cohomology is at the cross-road of both data analysis and statistical physics, and I aim to give some keys of its interpretation within those two fields, what could be  quoted as "passing the information between disciplines" in reference to Mezard's review \cite{Mezard2003}.\\     
\textbf{Unsupervised neural network learning:} the original work based on spin networks by Hopfield \cite{Hopfield1982} formalized fully recurrent networks as $n$ binary random variables ($N=2$). Ackley, Hinton and Sejnowski \cite{Ackley1985} followed up by imposing the Markov Field condition, allowing the introduction of conditional independence to handle network structures with hidden layers and hidden nodes. The result, the Boltzmann or Helmholtz machine \cite{Dayan1995}, relies on the maximum entropy or free-energy minimization principle, and originally on minimizing the relative entropy between the network and environmental states \cite{Ackley1985}. The informational approach of topological data analysis provides a direct probabilistic and statistical analysis of the structure of a dataset which allows to bridge the gap with neural networks analysis, and may be to go a step beyond their formalization in mathematic. Indeed, as proposed in \cite{Baudot2018a} and references therein (notably see the whole opus of Marcolli with application to linguistic), there are some reasons to believe that artificial or natural cognition could be homological by nature. 
Considering neuron as binary random variables provides in the present context an homologically constrained approach of those neural networks where the first input layer is represented by the marginal (single variable, degree 1  component) while hidden layers are associated to higher degrees. In a very naive sense higher cohomological degrees distinguish higher order patterns (or higher dimensional patterns in the simplicial case), just as receptive fields of neural networks recognize higher order features when going to higher depth-rank of neural layers as described in David Marr's original sketch \cite{Marr1982}, and now implemented efficiently in deep network structures.  
Here, in order to allow a direct interpretation of the information topology in standard terms of machine learning; I provide an energy functional interpretation of Mutual Information $I_k$ and of Total Correlation $G_k$ functions. Notably, the introduction of this multiplicity of "energy functions" broaden common unsupervised methods in neural network : instead of a single energy and associated gradient descent, Mutual informations provide a combinatorial family of analytically independent functions with independent gradients (see \cite{Baudot2019} for associated theorems). The scope is restricted to unsupervised learning here, but the supervised subcase with explicit homological information version of the back propagation chain rule, can be readily inferred from this and will be detailed with data analysis examples in future publication and illustrated here \cite{Baudot2019a}. Notably, the mathematical depth of back-propagation algorithm \cite{Kelley1960,Dreyfus1962,Rumelhart1986} comes from the fact that it implements the chain rule for derivation with respect to the parameters, allowing to learn in the class of differentiable functions. In the information cohomological reformulation, the chain rule becomes the chain rule of mutual-information that are coboundaries (commonly differential operator), parameters are provided by the label-teaching variables.  \\

\textbf{Statistical physics:} On the

\tableofcontents side of statistical physics, the generalization of binary neuronal-ising models to arbitrary multivalued variables with $N_i$ values corresponds to the generalization of spin models to Potts models \cite{Wu1982}. Such generalization is pertinent for biological systems in general, since coding and meaningful variables in biology are usually non-digital and only in quite exceptional cases binary. More interestingly, Mutual-Information negativity, a phenomenon known to provide the signature of frustrated states in glasses since the work of Matsuda \cite{Matsuda2001}, is related here, in the context of the more general conditional mutual information, to a kind of first order transition, yet seen topologically as the critical points of a simplicial complex. We propose to analyse the statistical interactions quantified by Mutual-Informations in full analogy with the Van Der Walls interaction model while leaving any homogeneous or mean-field assumptions on the variable. This observation makes coincide two phenomena, condensation in statistical physics and clustering in data science. We hence further suggest that the Mutual-Informations functions and homology should be peculiarly useful in the study of frustrated states and other related problems (see Mezard and Montanari \cite{Mezard2009a}), and for data points clustering.   \\

\textbf{Topological Data Analysis:} On the side of applied algebraic topology, the identification of the topological structures of dataset has motivated important researches following the development of persistent homology \cite{Lum2013,Epstein2011,Carlsson2009}. Combining in a single framework statistical and topological structures remains an active challenge of data analysis that already gave some interesting results \cite{Niyogi2011,Buchet2014}. Some recent works have proposed information theoretical approaches grounded on homology, defining persistent entropy \cite{Chintakunta2015,Merelli2015}, graph's topological entropy \cite{Tadic2016}, spectral entropy \cite{Maletic2012}, or multilevel Integration entropies \cite{Maletic2017}.
Although the present work is formally different and arose independently of persistence methods, it is possible to provide an intuitive non-rigorous interpretation in persistence terms. Most of the persistent methods consist in approximating the birth and death of the Betti's numbers of the \u{C}ech complex obtained by considering balls around each data point  while the radius of the ball grows. The \u{C}ech complex is given by the intersection of the balls, and for combinatorial computational reasons, most of the algorithm restrict to pairwise intersections giving the Vietoris-Ripps complex as an approximation of the \u{C}ech complex. Our method focuses on intersection of random variables rather than balls around data points: a theorem of Hu Kuo Ting \cite{Hu1962} (recalled in \cite{Baudot2019}) shows the equivalence of mutual information functions with set theoretic finite measurable functions endowed with the intersection operator, formalizing the naive Venn diagram usual interpretation of mutual informations. Hence, leaving the mathematical rigor to allow an interpretation of the current algorithm in the common language of persistence: I compute here an information theoretic analog of the \u{C}ech complex, and it is not excluded that such analogy can be made formal, and notably to establish some nerve theorem for information structures (see Oudot for review \cite{Oudot2015}). It turns out that "zero information intersections" is exactly equivalent to statistical independence (theorem 1.2 \cite{Baudot2019}).
Hence, in regard to current topological data analysis methods, the methods presented here provides an intrinsically probabilistic cohomological framework: the differential operators are fundamental maps in probability-information theory. As a consequence, no metric assumption is required apriori: in practice it is possible to compute the homology for example on position variables and-or on qualitative variable such as "nice" and "not nice". The present method is topological and avoids the introduction a priori of such a metric: rather, a family of Shannon's pseudometric emerges from the formalism as a first cohomological class (cf. section \ref{metric}): considering a symmetric action of conditioning, we obtain Shannon's metric parametrized by a scalar multiplicative constant. Notably, the notion of geodesic used in machine learning is replaced by the homotopical notion of path. On the data analysis side, it provides new algorithm and tools for Topological Data Analysis allowing to rank, detect clusters, functional modules and to make dimensionality reduction; all these classical tasks in data analysis have indeed a direct homological meaning. I propose to call the data analysis methods presented here, the Poincar\'{e}-Shannon machine, since it implements simplicial homology and information theory in a single framework, applied effectively to empirical data.

\section{Information cohomology}

This section provides a short bibliographical note on the inscription of information and probability theory within homological theories \ref{biblionote on homology}. We also recall the definition of information functions \ref{Information functions} and provide a short description of information cohomology computed in the low degrees \ref{Information structures and coboundaries}, such that the interpretation of entropy and mutual-informations within Hochschild cohomology appears straightforward and clear. There is no new result in this section, but I hope a more simple and helpful presentation for some researchers outside the field of topology, of what can be find in  \cite{Baudot2015a, Vigneaux2017, Vigneaux2019} that should be considered for more precise and detailed exposition. 

\subsection{A long march through information topology}\label{biblionote on homology}

From the mathematical point of view, a motivation of information topology is to capture the ambiguity theory of Galois, which is the essence of group theory or discrete symmetries (see Andr\'{e}'s reviews \cite{Andre2007b,Andre2008}), and Shannon's information uncertainty theory in a common framework, a path already paved by some results on information inequalities (see Yeung's results \cite{Yeung2003}) and in algebraic geometry. In the work of Cathelineau, \cite{Cathelineau1988}, entropy first appeared in the computation of the degree one homology of the group $SL(2,\mathbb{C})$  with coefficients in the adjoint action by choosing a pertinent definition of the derivative of the Bloch-Wigner dilogarithm. It could be shown that the functional equation with 5-terms of the dilogarithm implies the functional equation of entropy with 4-terms. Kontsevitch \cite{Kontsevitch1995} discovered that a finite truncated version of the logarithm appearing in cyclotomic studies also satisfied the functional equation of entropy, suggesting a higher degree generalization of information, analog to polylogarithm, and hence showing that the functional equation of entropy holds in p and 0 field characteristics. Elbaz-Vincent and Gangl used algebraic means to construct this information generalization which holds over finite fields \cite{Elbaz-Vincent2002}, and where information functions appear as derivations \cite{Elbaz-Vincent2015}. After entropy appeared in tropical and idempotent semi-ring analysis in the study of the extension of Witt semiring to the characteristic 1 limit \cite{Connes2009}, Marcolli and Thorngren developed thermodynamic semiring, and entropy operad that could be constructed as deformation of the tropical semiring \cite{Marcolli2011}. Introducing Rota-Baxter algebras, it allowed to derive a renormalization procedure \cite{Marcolli2014}. Baez, Fritz and Leinster defining the category of finite probability and using Fadeev axiomatization, could show that the only family of functions that has the functorial property is Shannon information loss \cite{Baez2011,Baez2014}. Boyom, basing his approach on information and Koszul geometry, developed a more geometrical view of statistical models that notably considers foliations in place of the random variables \cite{Boyom2016}.
Introducing a deformation theoretic framework, and chain complex of random variables, Drumond-Cole, Park and Terilla \cite{Drummond-Cole2015,Drummond-Cole2015a,Park2015}  could construct a homotopy probability theory for which the cumulants coincide with the morphisms of the homotopy algebras. A probabilistic framework, used here, was introduced in \cite{Baudot2015a}, and generalized to Tsallis entropies by Vigneaux \cite{Vigneaux2017,Vigneaux2019}.
The diversity of the formalisms employed in these independent but convergent approaches is astonishing. So, to the question what is information topology, it is only possible to answer that it is under development at the moment. The results of Catelineau, Elbaz-Vincent and Gangl inscribed information into the theory of motives, which according to Beilison's program is a mixed Hodge-Tate cohomology \cite{Beilinson1990}. All along the development of the application to data, following the cohomology developed by \cite{Baudot2015a,Vigneaux2017} on an explicit probabilistic basis, we aimed to preserve such a structure and unravel its expression in information theoretic terms. Moreover, following Aomoto's results \cite{Aomoto1982,Goncharov2005}, the actual conjecture \cite{Baudot2015a} is that the higher classes of information cohomology should be some kind of polylogarithmic k-form (k-differential volume that are symmetric and additive, and that correspond to the cocycle conditions for the cohomology of Lie groups \cite{Aomoto1982}). The following developments suggest that these higher information groups should be the families of functions satisfying the functional equations of k-independence $I_k=0$, a rather vague but intuitive view that can be tested in special cases.

\subsection{Information functions (definitions)} \label{Information functions}

The information functions used in \cite{Baudot2015a} and the present study were originally defined by Shannon \cite{Shannon1948} and Kullback \cite{Kullback1951} and further generalized and developed by Hu Kuo Ting \cite{Hu1962} and Yeung  \cite{Yeung2007} (see also McGill \cite{McGill1954}). These functions include entropy, noted $H_1=H(X;P)$, joint entropy, noted $H_k=H(X_1,...,X_k;P)$, mutual-information noted $I_2=I(X_1;X_2;P)$, multivariate k-mutual-information, noted $I_k=I(X_1;...;X_k;P)$ and the conditional entropy and mutual information, noted $Y.H_k=H(X_1,...,X_k|Y;P)$ and $Y.I_k=I(X_1;...;X_k|Y;P)$. 
The classical expression of these functions is the following (using $k=-1/\ln 2$, the usual bit unit):
\begin{itemize}
	\item The Shannon-Gibbs entropy of a single variable $X_j$ is defined by \cite{Shannon1948}: 
	\begin{equation}\label{singleentropy}
	H_1=H(X_{j};P_{X_j})=k\sum_{x \in [N_j] }p(x)\ln p(x)=k\sum_{i=1}^{N_j}p_i\ln p_i
	\end{equation}
	where $[N_j]=\{1,...,N_j\}$ denotes the  alphabet of $X_j$.

	\item The relative entropy or Kullback-Liebler divergence, which was also called "discrimination information" by Kullback \cite{Kullback1951}, is defined for two probability mass function $p(x)$ and $q(x)$ by: 
	\begin{equation}\label{relative entropy}
	\begin{split}
	D(p(x)||q(x)) & = D(X;p(x)||q(x))=k\sum_{x \in \mathscr{X}}p(x)\ln \frac{q(x)}{p(x)} \\
	& = H(X;p(x),q(x)) - H(X;p(x))
	\end{split}
	\end{equation}
	where $H(X;p(x),q(x))$ is the cross-entropy and $H(X;p(x))$ the Shannon entropy. It hence generates as a special case minus entropy, taking the deterministic constant probability $q(x)=1$. With the convention $k=-1/\ln 2$, $D(p(x)||q(x))$ is always positive or null.

	\item The joint entropy is defined for any joint-product of $k$ random variables  $(X_1,...,X_k)$ and for a probability joint-distribution $\mathbb{P}_{(X_1,...,X_k)}$  by \cite{Shannon1948}:
	\begin{equation} \label{jointentropy multiple}
	\begin{split}
	H_k & = H(X_{1},...,X_{k};P_{X_{1},...,X_{k}}) \\
	&  =  k\sum_{x_1,...,x_k\in [N_1\times...\times N_k]}^{N_1\times...\times N_k}p(x_1.....x_k)\ln p(x_1.....x_k) \\
	&  =  k\sum_{i,j,...,k}^{N_1,...,N_k}p_{\underbrace{ij...k}_\text{k indices}}\ln p_{ij...k}
	\end{split}
	\end{equation}
	where $[N_1\times...\times N_k]=\{1,...,N_j\times...\times N_k\}$ denotes the  alphabet of $(X_1,...,X_k)$.
	
	\item The mutual information of two variables $X_{1},X_{2}$ is defined as \cite{Shannon1948}: 
	\begin{equation}\label{mutual information}
	I(X_{1};X_{2};P_{X_{1},X_{2}})=k\sum_{x_1,x_2\in[N_1\times N_2]}^{N_1\times N_2}p(x_1.x_2)\ln \frac{p(x_1)p(x_2)}{p(x_1.x_2)}
	\end{equation}
	And it can be generalized to k-mutual-information (also called co-information) using the alternated sums given by equation \ref{Alternated sums of information}, as originally defined by McGill \cite{McGill1954} and Hu Kuo Ting \cite{Hu1962}, giving:
	\begin{equation}\label{n-mutual information}
	I_k=I(X_{1};...;X_{k};P)=k\sum_{x_1,...,x_k\in [N_1\times...\times N_k]}^{N_1\times...\times N_k}p(x_1.....x_k)\ln \frac{\prod_{I\subset [k];card(I)=i;i \ \text{odd}} p_I}{\prod_{I\subset [k];card(I)=i;i \ \text{even}} p_I} 
	\end{equation}
	For example, the 3-mutual information is the function: 
	\begin{equation}\label{3-mutual information}
	I_3=k\sum_{x_1,x_2,x_3\in[N_1\times N_2\times N_3]}^{N_1 \times N_2\times N_3}p(x_1.x_2.x_3)\ln \frac{p(x_1)p(x_2)p(x_3)p(x_1.x_2.x_3)}{p(x_1.x_2)p(x_1.x_3)p(x_2.x_3)}
	\end{equation}
	For $k\geq3$, $I_k$ can be negative \cite{Hu1962}.

	\item  The  total correlation introduced by Watanabe \cite{Watanabe1960}, called integration by Tononi and Edelman \cite{Tononi1998} or  multi-information by Studen\'{y} and  Vejnarova \cite{Studeny1999} and  Margolin and colleagues \cite{Margolin2010}, which we note $C_k(X_1;...X_k;P)$, is  defined by:
	\begin{equation}\label{total correlation}
	\begin{split}
	C_k &= C_k(X_1;...X_k;P)=\sum_{i=1}^k H(X_i) - H(X_1;...X_k)=\sum_{i=2}^{k}(-1)^{i}\sum_{I\subset [n];card(I)=i}I_i(X_I;P)\\
	& =k\sum_{x_1,...,x_k\in[N_1\times...\times N_k]}^{N_1 \times ...\times N_k}p(x_1....x_k)\ln \frac{p(x_1...x_k)}{p(x_1)...p(x_k)}
	\end{split}
	\end{equation}
	For two variables the total correlation is equal to the mutual-information ($C_2=I_2$). The total correlation has the nice property of being a relative entropy \ref{relative entropy} between marginal and joint-variable and hence to be always non-negative.  
	
	\item The conditional entropy of $X_{1}$ knowing (or given) $X_{2}$ is defined as \cite{Shannon1948}:
	\begin{multline}\label{conditionalentropy}
	X_{2}.H_{1}= H(X_{1}|X_{2};P)=k\sum_{x_1,x_2\in[N_1\times N_2]}^{N_1*N_2}p(x_1.x_2)\ln p_{x_2}(x_1) \\
	=  k \sum_{x_2\in\mathscr{X}_2}^{N_2}p(x_2). \left( \sum_{x_1\in\mathscr{X}_1}^{N_1}  p_{x_2}x_1  \ln p_{x_2}x_1 \right)
	\end{multline}
	Conditional joint-entropy, $X_3.H(X_1,X_2)$ or $(X_1,X_2).H(X_3)$, is defined analogously by replacing the marginal probabilities by the joint probabilities.  
	
	\item The conditional mutual information of two variables $X_{1},X_{2}$ knowing a third $X_3$ is defined as \cite{Shannon1948}: 
	\begin{equation}\label{conditional mutual information}
	X_3.I_2=I(X_{1};X_{2}|X_3;P)=k\sum_{x_1,x_2,x_3\in [N_1\times N_2\times N_3]}^{N_1\times N_2\times N_3}p(x_1.x_2.x_3)\ln \frac{p_{x_3}(x_1)p_{x_3}(x_2)}{p_{x_3}(x_1,x_2)} 
	\end{equation}
	Conditional mutual information generates all the preceding information functions as subcases, as shown by Yeung \cite{Yeung2007}. We have the theorem : if $X_3=\Omega$ then it gives the mutual information, if $X_2=X_1$ it gives conditional entropy, and if both conditions are satisfied, it gives entropy. Notably, we have $I_1=H_1$.
\end{itemize}

We now give the few information equalities and inequalities that are of central use in the homological framework, in the information diagrams and for the estimation of the informations from the data.\\
We have the chain rules (see \cite{Cover1991} for proofs): 
\begin{equation}\label{chain rule}
H(X_{1}; X_{2};P)=H(X_{1};P)+X_1.H(X_{2};P)=H(X_{2};P)+X_2.H(X_{1};P)
\end{equation}
\begin{equation}\label{chain rule infomut}
I(X_{1}; X_{2};P)=H(X_{1};P)-X_2.H(X_{1};P)=H(X_{2};P)-X_1.H(X_{2};P)
\end{equation}
That we can write more generally (where the hat denotes the omission of the variable):
\begin{equation}\label{chain rule gener}
H(X_1;...;\widehat{X_i} ;...;X_{k+1};P)  = H(X_1;...;X_{k+1};P) - (X_1;...;\widehat{X_i} ;...;X_{k+1}).H(X_i;P) 
\end{equation}
That we can write in short $H_{k+1} - H_k = (X_1,...X_k).H(X_{k+1})$
\begin{equation}\label{chain rule gene infomut}
I(X_1;...;\widehat{X_i} ;...;X_{k+1};P) = I(X_1;...;X_{k+1};P) + X_i.I(X_1;...;\widehat{X_i} ;...;X_{k+1};P) 
\end{equation}
That we can write in short $I_{k-1} - I_k = X_k.I_{k-1}$, generating the chain rule \ref{chain rule} as special case.

These two equations provide recurrence relationships that give an alternative formulation of the chain rules in terms of a chosen path on the lattice of information structures:
\begin{equation}\label{chain rule gener rec}
H_k=H(X_1,...,X_{k};P)  = \sum_{i=1}^{k} (X_1,...,X_{i-1}).H(X_i;P)
\end{equation}
where we assume $H(X_1;P)=X_0.H(X_1;P)$ and hence that $X_0$ is the greatest element $X_0=\Omega$.
\begin{equation}\label{chain rule gene infomut rec}
I_k=I(X_1;...;X_{k};P) = I(X_1) - \sum_{i=2}^{k} X_i.I(X_1;...;X_{i-1})
\end{equation}

We have the alternated sums or inclusion-exclusion rules \cite{Hu1962,Matsuda2001,Baudot2015a}: 
\begin{equation}\label{Alternated sums of entropy}
H_n(X_1,...,X_n;P)=\sum_{i=1}^{n}(-1)^{i-1}\sum_{I\subset [n];card(I)=i}I_i(X_I;P)
\end{equation}

\begin{equation}\label{Alternated sums of information}
I_n(X_1;...;X_n;P)=\sum_{i=1}^{n}(-1)^{i-1}\sum_{I\subset [n];card(I)=i}H_i(X_I;P)
\end{equation}
For example: $H_3(X_1,X_2,X_3)=I_1(X_1)+I_1(X_2)+I_1(X_3)-I_2(X_1;X_2)-I_2(X_1;X_3)-I_2(X_2;X_3)+I_3(X_1;X_2;X_3)$

The chain rule of mutual-information goes together with the following inequalities discovered by Matsuda \cite{Matsuda2001}. For all random variables $X_{1};..;X_{k}$ with associated joint probability distribution $P$ we have:
\begin{itemize} \label{infocond pos neg}
	\item  $X_{k}.I(X_{1};..;X_{k-1};P) \geq 0$ if and only if $I(X_{1};..;X_{k-1};P) \geq I(X_{1};..;X_{k};P)$ (in short: $I_{k-1} \geq I_k$ )
	\item $X_{k}.I(X_{1};..;X_{k-1};P) < 0$ if and only if $I(X_{1};..;X_{k-1};P) < I(X_{1};..;X_{k};P)$ (in short: $I_{k-1} < I_k$ )
\end{itemize}
that fully characterize the phenomenon of information negativity as an increasing or diverging sequence of mutual information.

\subsection{Information structures and coboundaries}\label{Information structures and coboundaries}

\begin{figure} [!h]
	\centering
	\includegraphics[height=5cm]{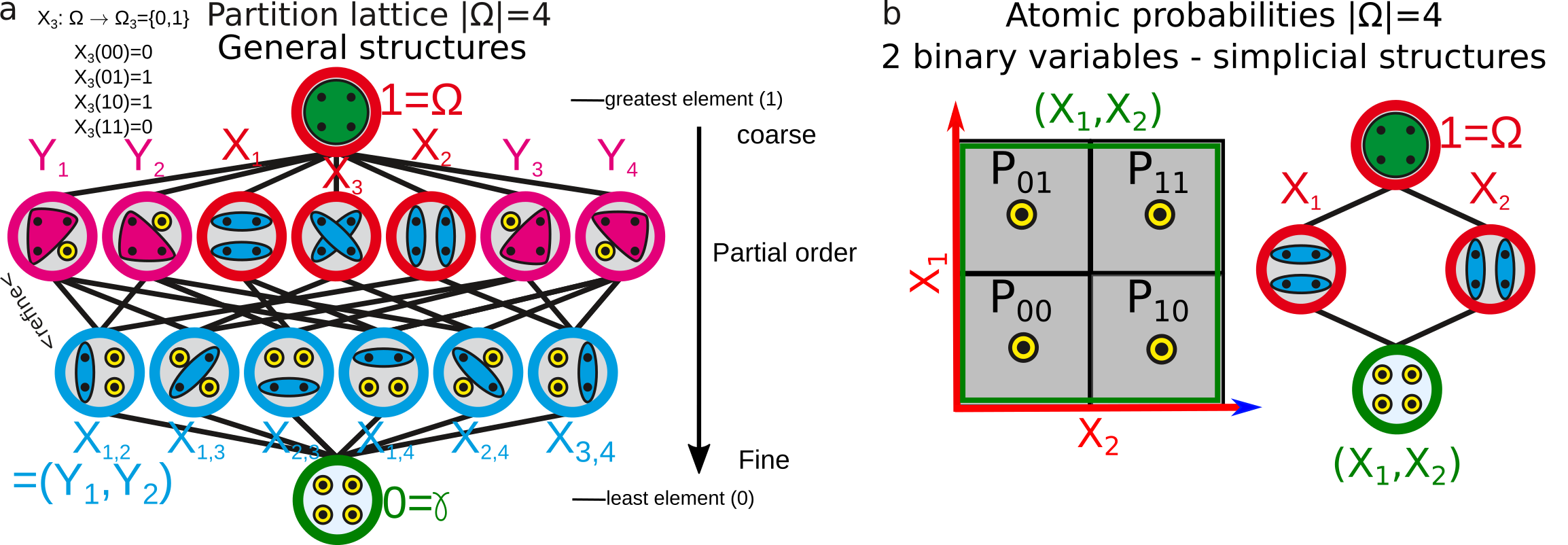}
	\caption{\textbf{Example of general and simplicial information structures. a,} Example of lattice of random variables (partitions): the lattice of partitions of atomic-elementary events for a sample space of 4 atomic elements $|\Omega|=4$ (for example two coins and $\Omega=\{00,01,10,11\}$), each element being denoted by a black dot in the circles representing the random variables. The joint operation of Random Variables noted $(X,Y)$ or $X\otimes Y$ of two partitions is the less fine partition $Z$ that is finer than $X$ and $Y$ ($Z$ divides $Y$ and $X$, or $Z$ is the greatest common divisor of $Y$ and $X$). It is represented by the coincidence of two edges of the lattices. The joint operation has an identity element noted $1=\Omega$ (that we will note 0 thereafter), with $X,1=X,\Omega=X$ and is idempotent $(X,X)=X^2=X$. The structure is a partially ordered set (poset) with a refinement relation.\textbf{ b,} Illustration of the simplicial structure (sublattice) used for the data analysis ( $|\Omega|=4$ as previously).}
	\label{figure_Supp_partition lattice}
\end{figure}

This section justifies the choice of functions and algorithm, the topological nature of the data analysis and the approximations we had to concede for the computation. In the general formulation of information cohomology, the random variables are partitions of the atomic probabilities of a finite probability space $(\Omega,\mathcal{B},P)$ (e.g. all their equivalence classes). The \textbf{Joint-Variable} $(X_1,X_2)$ is the less fine partition that is finer than $X_1$ and $X_2$; the whole lattice of partitions $\Pi$ \cite{Andrews1998} corresponds to the lattice of joint random variables \cite{Fresse2004,Baudot2015a}. Then, a general \textbf{information structure} is defined to be the triple $(\Omega,\Pi,P)$. A more modern and general expression in category theory and topos is given in \cite{Baudot2015a,Vigneaux2017}. $(X_1,...,X_k;P)$ designates the image law of the probability $P$ by the measurable function of joint variables $(X_1,...,X_k)$. Figure \ref{figure_Supp_partition lattice} gives a simple example of the lattice of partitions for 4 atomic probabilities, with the simplicial sublattice used for data analysis. Atomic probabilities are also illustrated in a Figure the associated paper \cite{Baudot2019}. 

On this general information structure, we consider the real module of all measurable functions $F(X_1,...,X_k;P)$, and the conditioning-expectation by $Y$ of measurable functions as the action of $Y$ on the functional module, noted $Y.F(X_1,...,X_k;P)$, such that it corresponds to the usual definition of conditional entropy (equ. \ref{conditionalentropy}). We define our complexes of measurable functions of random variables $X^k=F(X_1,...,X_k;P)$,  and the cochain complexes $(X^k,\partial^k)$ as :
\begin{equation*}
0 \xrightarrow{} X^0 \xrightarrow{\partial^0} X^1 \xrightarrow{\partial^1} X^{2} \xrightarrow{\partial^2} ... X^{k-1}\xrightarrow{\partial^{k-1}} X^{k}
\end{equation*}
where $\partial^k$ is the left action co-boundary that Hochschild proposed for associative and ring structures \cite{Hochschild1945}. A similar construction of random variable complex was given by Drumond-Cole, Park and Terilla \cite{Drummond-Cole2015,Drummond-Cole2015a}. We consider also the two other directly related cohomologies that are defined by considering a trivial left action \cite{Baudot2015a} and a symmetric (left and right) action  \cite{Gerstenhaber1987,Weibel1995,Kassel2004} of conditioning:  
\begin{itemize}
	\item The left action Hochschild-information coboundary and cohomology (with trivial right action):
	\begin{equation}
	\begin{split}
	(\partial^k)F(X_1;X_2;...;X_{k+1};P)={} & X_1.F(X_2;...;X_{k+1};P)\\
	& +\sum_{i=1}^{k}(-1)^{i} F(X_1; X_2;...;(X_i,X_{i+1});...;X_{k+1};P)\\
	& + (-1)^{k+1} F(X_1;...;X_{k};P)
	\end{split}
	\end{equation} 
	This coboundary, with a trivial right action, is the usual coboundary of Galois cohomology (\cite{Tate1991}, p.2), and in general it is the coboundary of homological algebra obtained by Cartan and Eilenberg \cite{Cartan1956} and MacLane \cite{MacLane1975} (non homogenous bar complex). 	
	\item The "topological-trivial" Hochschild-information  coboundary and cohomology: considering a trivial left action in the preceding setting , e.g. $X_1.F(X_2;...;X_{k+1})=F(X_2;...;X_{k+1})$. It is the subset of the preceding case, which is invariant under the action of conditioning. We obtain the topological coboundary $(\partial_t^k)$ \cite{Baudot2015a}:
	\begin{equation}
	\begin{split}
	(\partial_t^k)F(X_1;X_2;...;X_{k+1};P)={} & F(X_2;...;X_{k+1};P)\\
	& +\sum_{i=1}^{k}(-1)^{i} F(X_1; X_2;...;(X_i,X_{i+1});...;X_{k+1};P)\\
	& + (-1)^{k+1} F(X_1;...;X_{k};P)
	\end{split}
	\end{equation} 
	\item The symmetric Hochschild-information coboundary and cohomology: as introduced by Gerstenhaber and Shack \cite{Gerstenhaber1987}, Kassel  \cite{Kassel2004} (p.13) and Weibel \cite{Weibel1995} (chap.9), we consider a symmetric (left and right) action of conditioning, that is  $X_1.F(X_2;...;X_{k+1})=F(X_2;...;X_{k+1}).X_1$. The left action module is essentially the same as considering a symmetric action bimodule \cite{Gerstenhaber1987,Kassel2004,Weibel1995}. We hence obtain the following symmetric coboundary $(\partial_{\ast}^k)$:
	\begin{equation}
	\begin{split}
	(\partial_{\ast}^k)F(X_1;X_2;...;X_{k+1};P)={} & X_1.F(X_2;...;X_{k+1};P)\\
	& +\sum_{i=1}^{k}(-1)^{i} F(X_1; X_2;...;(X_i,X_{i+1});...;X_{k+1};P)\\
	& + (-1)^{k+1} X_{k+1}.F(X_1;...;X_{k};P)
	\end{split}
	\end{equation} 
\end{itemize}
Based on these definitions, Baudot and Bennequin \cite{Baudot2015a} computed the first homology class in the left action Hochschild-information cohomology case and the coboundaries in higher degrees. We introduce here the symmetric case, and detail the higher degree cases by direct specialization of the co-boundaries formulas, such that it appears that information functions and chain rules are homological by nature. 
For notation clarity, we omit the probability in the writing of the functions, and  when specifically stated replace their notation $F$ by their usual corresponding informational function notation $H,I$.
\subsubsection{first degree (k=1)}
For the first degree $k=1$, we have the following results:
\begin{itemize}
	\item The left 1-co-boundary is $(\partial^1)F(X_1;X_2)= X_1.F(X_2)-F(X_1,X_2)+ F(X_1)$. The 1-cocycle condition  $(\partial^1)F(X_1;X_2)=0$ gives  $F(X_1,X_2)= F(X_1)+X_1.F(X_2)$, which is the chain rule of information shown in equation \ref{chain rule}. Then, following Kendall \cite{Kendall1964} and Lee \cite{Lee1964}, it is possible to recover the functional equation of information and to characterize uniquely, up to the arbitrary multiplicative constant $k$, the entropy (equation \ref{singleentropy}) as the first class of cohomology \cite{Baudot2015a,Vigneaux2017}. This main theorem allows us to obtain the other information functions in what follows. Marcolli and Thorngren \cite{Marcolli2011}, Leinster, Fritz and Baez \cite{Baez2014,Baez2011} obtained independently an analog result using measure-preserving function and characteristic one Witt construction, respectively. In these various theoretical settings, this result extends to relative entropy \cite{Marcolli2011,Baez2014,Baudot2015a}, and Tsallis entropies \cite{Marcolli2011,Vigneaux2017}.
	
	\item  The topological 1-coboundary $(\partial_t^1)$ is $(\partial_t^1)F(X_1;X_2)= F(X_2)-F(X_1,X_2)+F(X_1)$, which corresponds to the definition of mutual information $(\partial_t^1)F(X_1;X_2)=I(X_1;X_2)= H(X_1)+H(X_2)-H(X_1,X_2)$, and hence $I_2$ is a topological 1-coboundary.  
	
	\item  The symmetric 1-coboundary $(\partial_{\ast}^1)$ is $(\partial_{\ast}^1)F(X_1;X_2)= X_1.F(X_2)-F(X_1,X_2)+X_2.F(X_1)$, which corresponds to the negative of the pairwise mutual information $(\partial_{\ast}^1)F(X_1;X_2)= X_2.H(X_1)+X_1.H(X_2)-H(X_1,X_2)=-I(X_1;X_2)$, and hence $-I_2$ is a symmetric 1-coboundary. Moreover, the 1-cocycle condition $(\partial_{\ast}^1)F(X_1;X_2)=0 $ characterizes functions satisfying $F(X_1,X_2)=X_2.F(X_1)+X_1.F(X_2)$, which corresponds to the information pseudo-metric discovered by Shannon \cite{Shannon1953}, Rajski  \cite{Rajski1961}, Zurek  \cite{Zurek1989} and Bennett \cite{Bennett1998}, and has further been applied for hierarchical clustering and finding categories in data by Kraskov and Grassberger \cite{Kraskov2009}: \label{metric} $H(X_1\bigtriangleup X_2)=X_2.H(X_1)+X_1.H(X_2)= H(X_1,X_2)-I(X_1;X_2)$. Therefore, up to an arbitrary scalar multiplicative constant $k$, the information pseudo-metric  $H(X_1\bigtriangleup X_2)$ is the first class of symmetric cohomology. This pseudo metric is represented in the Figure \label{Figure_glued_landscape}. It generalizes to pseudo k-volumes that we define by $V_k=H_k-I_k$ (particularly interesting symmetric functions computed by the provided software). 	
\end{itemize}

\subsubsection{Second degree (k=2)}

For the second degree $k=2$, we have the following results:
\begin{itemize}
	\item The left 2-co-boundary is $\partial^{2}F(X_1; X_2;X_3)=X_1.F(X_2;X_3)-F((X_1,X_2);X_3)+F(X_1;(X_2,X_3))-F(X_1;X_2)$, which corresponds to minus the 3-mutual information $\partial^{2}F(X_1; X_2;X_3)=X_1.I(X_2;X_3)-I((X_1,X_2);X_3)+I(X_1;(X_2,X_3))-I(X_1;X_2)=-I(X_1;X_2;X_3)$, and hence $-I_3$ is left 2-coboundary.
	
	\item  The topological 2-coboundary is $(\partial_t^2)F(X_1;X_2;X_3)= F(X_2;X_3)-F((X_1,X_2);X_3)+F(X_1;(X_2,X_3))-F(X_1;X_2)$, which corresponds in information to $\partial_t^{2}F(X_1; X_2;X_3)= I(X_2;X_3)-I((X_1,X_2);X_3)+I(X_1;(X_2,X_3))-I(X_1;X_2)
	=0$, and hence the topological 2-coboundary is always null-trivial.  
	
	\item  The symmetric 2-coboundary is $(\partial_{\ast}^2)F(X_1;X_2;X_3)= X_1.F(X_2;X_3)-F((X_1,X_2);X_3)+F(X_1;(X_2,X_3))-X_3.F(X_1;X_2)$, which corresponds in information to $\partial_\ast^{2}F(X_1; X_2;X_3)= X_1.I(X_2;X_3)-I((X_1,X_2);X_3)+I(X_1;(X_2,X_3))-X_3.I(X_1;X_2)	=0$, and hence the symmetric 2-coboundary is always null-trivial.  	
\end{itemize}

\subsubsection{Third degree (k=3)}

For the third degree $k=3$, we have the following results:
\begin{itemize}
	\item The left 3-co-boundary is $\partial^{3}F(X_1; X_2;X_3;X_4)=X_1.F(X_2;X_3;X_4)-F((X_1,X_2);X_3;X_4)+F(X_1;(X_2,X_3);X_4)-F(X_1;X_2;(X_3,X_4))+F(X_1;X_2;X_3)$, which corresponds in information to  $\partial^{3}F(X_1; X_2;X_3;X_4)=X_1.I(X_2;X_3;X_4)-I((X_1,X_2);X_3;X_4)+I(X_1;(X_2,X_3);X_4)-I(X_1;X_2;(X_3,X_4))+I(X_1;X_2;X_3)=0$, and hence the left 3-coboundary is always null-trivial.  
	
	\item  The topological 3-coboundary is $\partial_t^{3}F(X_1;X_2;X_3;X_4)=F(X_2;X_3;X_4)-F((X_1,X_2);X_3;X_4)+F(X_1;(X_2,X_3);X_4)-F(X_1;X_2;(X_3,X_4))+F(X_1;X_2;X_3)$, which corresponds in information to $\partial_t^{3}F(X_1; X_2;X_3;X_4)=I(X_2;X_3;X_4)-I((X_1,X_2);X_3;X_4)+I(X_1;(X_2,X_3);X_4)-I(X_1;X_2;(X_3,X_4))+I(X_1;X_2;X_3)=I(X_1;X_2;X_3;X_4)$, and hence $I_4$ is a topological 3-coboundary.
	
	\item  The symmetric 3-coboundary is $(\partial_{\ast}^3)F(X_1;X_2;X_3;X_4)=X_1.F(X_2;X_3;X_4)-F((X_1,X_2);X_3;X_4)+F(X_1;(X_2,X_3);X_4)-F(X_1;X_2;(X_3,X_4))+X_4.F(X_1;X_2;X_3)$, which corresponds in information to $\partial_{\ast}^{3}F(X_1; X_2;X_3;X_4)=X_1.I(X_2;X_3;X_4)-I((X_1,X_2);X_3;X_4)+I(X_1;(X_2,X_3);X_4)-I(X_1;X_2;(X_3,X_4))+X_4.I(X_1;X_2;X_3)=-I(X_1;X_2;X_3;X_4)$, and hence $-I_4$ is a symmetric 3-coboundary.	
\end{itemize}

\subsubsection{Higher degrees}

For $k=4$, we obtain  $\partial^{4}F(X_1; X_2;X_3;X_4;X_5)=-I_5$, and $\partial_t^{5}F(X_1; X_2;X_3;X_4;X_5)=0$ , and $\partial_\ast^{5}F(X_1; X_2;X_3;X_4;X_5)=0$. For arbitrary $k$, the symmetric coboundaries are just the opposite of the topological coboundaries $\partial_{t}^k=-\partial_{\ast}^k$. It is possible to generalize to arbitrary degrees  \cite{Baudot2015a} by remarking that we have:
\begin{itemize}
	\item For even degrees $2k$: we have $I_{2k}=-\partial_{t}I_{2k-1}$  and then $I_{2k}=\partial_{t}\partial \partial_{t}...\partial\partial_{t}H$ with $2k-1$ boundary terms. 
	In conclusion, we have:
	\begin{equation}
	\partial^{2k}F=-I_{2k+1} ~ \text{ and }  ~ \partial_{\ast}^{2k}F=-\partial_{t}^{2k}F=0
	\end{equation} 
	\item For odd degrees $2k+1$: $I_{2k+1}=-\partial I_{2k-1}$  and then $I_{2k+1}=-\partial\partial_{t} \partial...\partial\partial_{t}H$ with $2k$ boundary terms.
	In conclusion, we have:
	\begin{equation}
	\partial^{2k-1}F=0 ~ \text{ and }  ~ \partial_{\ast}^{2k-1}F=-\partial_{t}^{2k}F=-I_{2k} 
	\end{equation} 
\end{itemize}

In \cite{Baudot2018,Baudot2019}(theorem 1.2), we show that the mutual independence of n variables is equivalent to the vanishing of all $I_k$ functions for all $2\leq k \leq n$. As a probabilistic interpretation and conclusion, the information cohomology hence quantifies statistical dependences at all degrees and  the obstruction to factorization. Moreover, $k$-independence coincides with cocycles. We therefore expect that the higher cocycles of information, conjectured to be polylogarithmic forms \cite{Baudot2015a,Elbaz-Vincent2015,Elbaz-Vincent2002}, are characterized by the functional equations $I_k=0$, and quantify statistical k-independence.

\section{Simplicial information cohomology}

\subsection{Simplicial substructures of information}

The general information structure, relying on the information functions defined on the whole lattice of partitions, encompasses all possible statistical dependences and relations, since by definition it considers all possible equivalent classes on a probability space. One could hence expect this general structure to provide the promising theoretical framework for classification tasks on data, and this is probably true in theory. However, this general case hardly allows any interesting computational investigation as it implies an exhaustive exploration of computational complexity following Bell's combinatoric in $\mathcal{O}(exp(exp(N^n))$) for $n$ $N$-ary variables. This fact was already remarked in the  study of aggregation for Artificial Intelligence by Lamarche-Perrin and colleagues \cite{Lamarche-Perrin2013}. At each order $k$, the number of $k$-joint-entropy and $k$-mutual-information to evaluate is given by Stirling numbers  of the second kind $S(n,k)$ that sum to Bell number $B_{n}$, $B_{n}=\sum_{k=0}^{n}S(n,k)$. For example, considering 16 variables that can take 8 values each, we have $8^{16}=2^{48} \approx 3.10^{14}$ atomic probabilities and the partition lattice of variables exhibits around $e^{e^{2^{48}}-1}\geq2^{200} $ elements to compute. Such computational reef can be decreased by considering the sample-size $m$, which is the number of trials, repetitions or points that is used to effectively estimate the empirical probability. It restricts the computation to $\mathcal{O}(exp(exp(m))$, which remains insurmountable in practice with our current classical Turing machines.
To circumvent this computational barrier, data analysis is developed on the simplest and oldest subcase of Hochschild cohomology: the simplicial cohomology, which we hence call the simplicial information cohomology and structure, and which corresponds to a subcase of cohomology and structure introduced previously (see Figure \ref{figure_Supp_partition lattice}b.).  It corresponds to the example 1 and 4 in \cite{Baudot2015a}. For simplicity, we note also the simplicial information structure $(\Omega,\Delta^n,P)$, $\Delta^n=(X_1,...,X_n;P)$, as we will not come back to the general setting.  Joint $(X_1,X_2)$ and meet $(X_1;X_2)$ operations on random variables are the usual joint-union and meet-intersection of Boolean algebra and define two opposite-dual monoids, generating freely the lattice of all subsets and its dual. The combinatorics of the simplicial information structure follow binomial coefficients and, for each degree $k$ in an information structure of $n$ variables, we have $\binom{n}{k}=\frac{n!}{k!(n-k!)}$ elements that are in one to one correspondence with the $k$-faces (the $k$-tuples) of the $n$-simplex of random variables (or its barycentric subdivisions). 
It is a (simplicial) substructure of the general structure since any finite lattice is a sub-lattice of the partition lattice \cite{Pudlak1980}. This lattice embedding and the fact that simplicial cohomology is a special case of Hochschild cohomology can also be inferred directly from their coboundary expression and has been explicitly formalized in homology: notably, Gerstenhaber and Shack showed that a functor, noted $\varSigma \mapsto k_\varSigma !$, induces an isomorphism between simplicial and Hochschild cohomology  $H^{\bullet}(\varSigma,k)\cong H^{\bullet}(k_\varSigma !,k_\varSigma !)$ \cite{Gerstenhaber1983}. A simplicial complex $X^k=F(X_1,...,X_k;P)$ of measurable functions is any subcomplex of this simplex  $\Delta^n$ with $k\leq n$, and any simplicial complex can be realized as a subcomplex of a simplex (see Steenrod \cite{Steenrod1947} p.296). The information landscapes presented in Figure \ref{figure_Supp_infolandscape} illustrate an example of such a lattice/information structure. Moreover in this ordinary homological structure, the degree obviously coincides with the dimension of the data space (the data space is in general $\mathbb{R}^n$, the space of "co-ordinate" values of the variables). 
This homological (algebraic, geometric and combinatorial) restriction to the simplicial subcase can have some important statistical consequences. In practice, whereas the consideration of the partition lattice ensured that no reasonable (up to logical equivalence) statistical dependences could be missed (since all the possible equivalence classes on the atomic probabilities were considered), the monoidal simplicial structure unavoidably misses some possible statistical dependences as shown and exemplified by James and Crutchfield \cite{James2017}.

\subsection{Topological Self and Free Energy of k-body interacting system - Poincar\'{e}-Shannon Machine} \label{Free information energy}

\subsubsection{Topological Self and Free Energy of k-body interacting system}\label{Free information_n-body interacting_system}

\begin{figure} [!h]
	\centering
	\includegraphics[height=4cm]{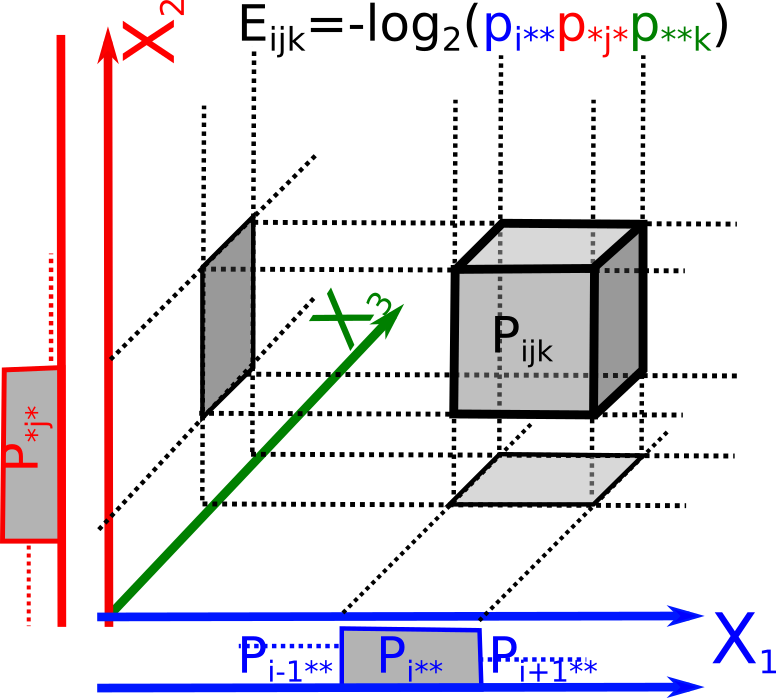}
	\caption{\textbf{Elementary energy as logarithm of locally Euclidean probability volumes}. Example of an elementary energy $E_{ijk}$ associated to a probability  $p_{ijk}$ ($n=3$ variables). The histograms of the marginal distributions of each variable are plotted beside the axes.}  
	\label{figure_Supp_elementary_energy}
\end{figure}

The basic idea behind the development of topological quantum field theories \cite{Atiyah1988,Witten1988,Schwarz2000} was to define the action and  energy functionals on a purely topological ground, independently of any metric assumptions, and to derive from this the correlation functions or partition functions. Here, in an elementary model for applied purposes, we define, in the special case of classical and discrete probability, the k-mutual-information $I_k$ (that generalize the correlation functions to nonlinear relation \cite{Reshef2011}), as the contribution of the k-body interactions to the energy functional. Some further observations support such a definition: i) as stated in \cite{Baudot2015a} (Th.D), the signed mutual-informations $(-1)^k I_k$ defining energy are sub-harmonic, a kind of weak convexity ii) in the next sections, we define the paths of information and show that they are equivalent to the discrete symmetry group iii) from the empirical point of view, Figure \ref{figure_meanHk_Ik} shows that these energy functionals estimated on real data behave as expected for usual k-body homogeneous formalism such as Van-Der-Walls model, or more refined Density Functional Theory (DFT) \cite{Hohenberg1964,Kohn1965}. These definitions, given in the context of simplicial structures, generalize to the case of partitions lattice, and altogether provide the  usual thermodynamical and machine-learning expressions and interpretation of mutual-information quantities: some new methods free of metric assumptions. There are two qualitatively and formally different components in the $I_k$ :
\begin{itemize}
	\item \textbf{Self-internal information energy (definition):} for $k=1$, $I_1$ and their sum in an information structure expressed in equation \ref{Alternated sums of entropy}, namely  $\sum_{T\subset [n];card(T)=1}I_1(X_T;P)$, are  a self-interaction component, since it sums over marginal information-entropy $I_1(X_i)=H_1(X_i)$. We call the first dimension mutual information component $U(X_1,...,X_n;P_N)$ the self information or internal information energy, in analogy to usual statistical physics and notably DFT:
	\begin{equation}\label{kinetic}
	U(X_1,...,X_n;P_N)=\sum_{i=1}^n I_1(X_i;P_N)
	\end{equation} 
	Note that in the present context, which is discrete and where the interactions do not depend on a metric, the self-interaction does not diverge, which is a usual problem with metric continuous formalism and was the original motivation for regularization and renormalization infinite corrections, considered by Feynman and Dirac as the  mathematical default of the formalism \cite{Feynman1985,Dirac1978}. 
	
	\item \textbf{k-free-energy and total-free-energy (definition)}: for $k\geq2$,  $(-1)^k I_k$ and their sum in an information structure (equation \ref{Alternated sums of entropy}) quantify the contribution of the k-body interactions. We call the k$^{\text{th}}$ dimension mutual information component $(-1)^k I_k$, given in equation \ref{n-mutual information}, the k-free-information-energy. We call the (cumulative) sum over dimensions of these k-free-information-energies starting at pairwise interactions (dimension 2), the total n-free-information-energy, and note it $G(X_1,...,X_n;P_N)$:
	\begin{equation}\label{free}
	G(X_1,...,X_n;P_N)=\sum_{i=2}^{n}(-1)^{i-1}\sum_{I\subset [n];card(I)=i}I_i(X_I;P_N)=C_n(X_1;...X_n;P_N)
	\end{equation}
	The total free-energy is the total correlation (equation \ref{total correlation}) introduced by Watanabe in 1960 \cite{Watanabe1960} that quantifies statistical dependence in the work of Studen\'{y} and Vejnarova \cite{Studeny1999} and  Margolin and colleagues \cite{Margolin2010}, and among other examples consciousness in the work of Tononi and Edelman \cite{Tononi1998}.  In agreement with the results of Baez and Pollard in their study of biological dynamics using out-of-equilibrium formalism \cite{Baez2016a}, and the appendix of the companion paper on Bayes Free Energy \cite{Baudot2019} the total free-energy is a relative entropy. The consideration that free energy is the peculiar case of total correlation within the set of relative entropies, accounts for the fact that the free energy shall be a symmetric function of the variables associated to the various bodies, e.g $f(X;Y)=f(Y;X)$ in the pairwise interaction case.  
	Moreover, whereas $I_k$ energy component can be negative, the $G_k$ total energy component is always non-negative. Each $(-1)^k I_k$ term in the free energy can be understood as a free energy correction accounting for the k-body interactions.	
\end{itemize}
Entropy is given by the alternated sums of information (equation \ref{Alternated sums of entropy}), which then read as the usual isotherm thermodynamic relation:
\begin{equation}\label{global_free}
H_n(X_1,...,X_n;P_N)=U(X_1,...,X_n;P_N)-G(X_1,...,X_n;P_N)
\end{equation}
This information theoretic formulation of thermodynamic relation follows Jaynes \cite{Jaynes1957,Jaynes1957a}, Landauer \cite{Landauer1961}, Wheeler \cite{Wheeler1983a}, and Bennett's\cite{Bennett2003} original work, and is general in the sense that it is finite and discrete, and holds independently of the assumption of the system being in equilibrium or not, i.e. for whatever finite probability. In more probabilistic terms, it does not assume that the variables are identically distributed, a condition that is required for the application of classical central limit theorems (CLT) to obtain the normal distributions in the asymptotic limit \cite{VonBahr1967}.  
In the special case where one postulates that the probability follows the equilibrium Gibbs distribution, which is also the maximum entropy distribution   \cite{Ebrahimi2008,Conrad2005}, the expression of the joint-entropy ($k=-1/\ln2$) allows to recover the equilibrium fundamental relation, as usually achieved in statistical physics (see Adami and Cerf \cite{Adami1999} and  Kapranov \cite{Kapranov2011} for more details). Explicitly, let's consider the Gibb's distribution:
\begin{equation}\label{Gibb's distribution }
p(X_1=x_1,...,X_n=x_n)=p_{\underbrace{ij...n}_\text{n indices}}=\frac{1}{Z}e^{-\beta E_{ij...n}/k_BT}
\end{equation}
where $E_{ij...n}$ is the energy of the elementary-atomic probability $p_{ij...n}$, $k_B$ is Boltzmann constant, $T$  the temperature and $Z=\sum_{i,j,..,n}^{N_i.N_j...N_n} e^{- E_{ij...n}/k_BT}$ is the partition function, such that $\sum_{i,j,..,n}^{N_i.N_j...N_n}  p_{ij...n}=1$. Since $H(X_1,...,X_n)=k\sum_{i,j,..,n}^{N_i.N_j...N_n} p_{ij...n} \ln p_{ij...n}$, equals the thermodynamic entropy function $S$  up to the arbitrary Landauer constant factor $k_B\ln2$, $S=k_B\ln2 H(X_1,...,X_n)$, the entropy for Gibbs distribution gives:
\begin{equation}\label{ entropy for Gibbs distribution}
H(X_1,...,X_n)/k= \sum_{i,j,..,n}^{N_1.N_2...N_n} p_{ij...n}E_{ij...n}/k_BT+ \sum_{i,j,..,n}^{N_1.N_2...N_n} p_{ij...n} \ln Z = (\langle E\rangle - G)/k_BT
\end{equation}
, which gives the expected thermodynamical relation: 
\begin{equation}\label{thermodynamical relation}
k_BT\ln2. H(X_1,...,X_n)=\langle E\rangle - G=U-G
\end{equation} 
, where $G$ is the free-energy $G=- k_BT\ln Z $. \\
In the general case of arbitrary random variables (not necessarily iid) and discrete probability space, the identification of marginal informations with internal energy:
\begin{equation}
\sum_{k=1}^n H(X_k)=\sum_{i,j,..,n}^{N_1.N_2...N_n} p_{ij...n}E_{ij...n}
\end{equation}
implies by direct algebraic calculus that:
\begin{equation}\label{Elementary_energy}
\sum_{i,j,..,n}^{N_1.N_2...N_n} p_{ij...n}E_{ij...n}=-\sum_{i,j,..,n}^{N_1.N_2...N_n} p_{ij...n} \ln \left( \prod_{k=i}^n p_{\bullet\bullet...k...\bullet}\right)
\end{equation} 
, where  the marginal probability $p_{\bullet\bullet...k...\bullet}$ is the sum over all probabilities for which $X_k=x_k$. It is hence tempting to identify the elementary-atomic energies $E_{ij...n}$ with the  elementary marginal informations $\ln p_{\bullet\bullet...k...\bullet}$. This is achieved uniquely by considering that such an elementary energy function must satisfy the additivity axiom (extensivity):  ($E(X_i=x_i,Xj=x_j)=E_{i,j}=E_{ij}=E_{i}+E_{j}$), 
which is the functional equation of the logarithm. The original proof goes back at least to Kepler, an elementary version was given by Erdos \cite{Erdoes1946}, and in Information theory terms can be found in the proofs of uniqueness of "single event information function" by Aczel and Darokzy (\cite{Aczel1975}, p.3). It establishes the following proposition: 

\begin{thm}
	Given a simplicial information structure, the elementary energies satisfying the extensivity axiom are the functions:  
	\begin{equation}\label{Elementary_energy_bis}
	E_{ij...n}=k\sum_{k=i}^n \ln p_{\bullet\bullet...k...\bullet}
	\end{equation} 
	, where $k$ is an arbitrary constant settled to $k=-1/\ln2$ for units in bit. 
\end{thm}
The geometric meaning of these elementary energies as log of marginal elementary probability volumes (locally Euclidean) is illustrated in Figure \ref{figure_Supp_elementary_energy} and further underlines that $I_{k,k\geq 2}$  are volume corrections accounting for the statistical dependences among marginal variables.\\
\textbf{Examples:} i) in the example of 3 binary random variables ($n=3,N_1=N_2=N_3=2$, three variables of Bernoulli) illustrated in the Figure of the associated paper \cite{Baudot2019}, we have $E_{000}=-\ln(p_{0\bullet\bullet}p_{\bullet0\bullet}p_{\bullet \bullet0})$, $E_{000}=-\ln(p_{000}+p_{010}+p_{001}+p_{011})-\ln(p_{000}+p_{100}+p_{001}+p_{101})-\ln(p_{000}+p_{100}+p_{010}+p_{110})$ and in the  configuration of negative-entangled-Borromean information of the Figure of the associated paper \cite{Baudot2019}, we obtain $E_{000}= 3$ in bit units, and similarly $E_{001}=E_{010}=E_{011}=E_{101}=E_{110}=E_{111}= 3$, and we hence recover $U=\sum_{i,j,k}^{8} p_{ijk}E_{ijk}=\sum_{i=1}^3H(X_i)=3$ bits. Note that the limit $0\ln0 \sim 0$ avoids singularity of elementary energies.\\
ii) in the special case of identically distributed variables, $p_{\bullet\bullet...k...\bullet}=p_{\bullet...j...\bullet\bullet}$, we have $E_{ij...n}=nk \ln p_{\bullet\bullet...k...\bullet}$ and hence the  marginal Gibbs distribution:  $p_{\bullet\bullet...k...\bullet}=e^{\frac{E_{ij...n}}{nk}}$ \label{Elementary_energyiid}. \\
iii) for independent identically distributed variable (non-interacting), we have $G_n=0$, and hence:\\
\begin{equation}
H_n(X_1,...,X_n;P_N)=U(X_1,...,X_n;P_N)=nH(X_i)
\end{equation}  
iv) considering the variables to be the $6n$ variables  of the phase space, with one variable of position and one variable of momentum per body (noted $(X^1_k,X^2_k,X^3_k,P^1_k,P^2_k,P^3_k)$ for the kth body), it is possible to re-express the semi-classical formalism according to which the entropy formulation is (Landau and Lifshitz \cite{Landau1969},p.22):
\begin{equation}
H_{6n}(X^1_1,X^2_1,X^3_1,P^1_1,P^2_1,P^3_1,...,P^3_n;P_N)=\log\left(\frac{\Delta X \Delta P}{(2\pi \hbar)^{6n}}\right)
\end{equation} 
It is achieved by identifying the internal and free energy as following:
\begin{equation}\label{semi-classical}
\langle E\rangle=-6n \log (2\pi \hbar)\\
\end{equation}
\begin{equation}\label{semi-classical2}
G=- \log (\Delta X \Delta P)
\end{equation}
This identifies the elementary volumes/probabilities with the Planck constant, the quantum of action (the consistency in the units is realized in section \ref{2nd law} by the introduction of time). The quantum of action can be illustrated by considering in the Figure \ref{figure_Supp_elementary_energy} that it is the surface of the square/rectangle for two conjugate variables (considered as position and momentum).  In this setting, $\Delta X \Delta P$  quantifies the non-extensivity of the volume in the phase-space due to interactions, or in other words, the mutual-informations account for the consideration of the dependence of the subsystems considered as opened and exchanging energy. As noted by Baez and Pollard, the relative entropy provides a quantitative measure of how far from equilibrium the whole system is \cite{Baez2016a}. The basic principle of such expression of information theory in physics is known at least since Jaynes's work \cite{Jaynes1957a,Jaynes1963}.  \\

As a conclusion, information topology applies, without imposing metric or symplectic or contact structures, to physical formalism of n-body interacting systems relying on empirical measures. Considering the $3n$ or $6n$ dimensions (degrees of freedom) of a configuration or a phase space as random variables, it is possible to recover the (semi) classical statistical physics formalism. It is also interesting to discuss the status of the analog of the temperature variable in the present formalism which is played by the graining, which is the size $N_i$ of the alphabet of a variable $X_i$. In usual thermodynamic we have $H(X^{n};P_N)=T.S(X^{n})$, and to stay consistent, temperature shall be a functional inverse of the graining $N$, lowest temperature being the finest grain (large $N$), highest temperature being the coarsest graining (small $N$).

\subsection{k-Entropy and k-Information landscapes} 

\textbf{Information landscapes (definition):}\\ 
The information landscapes are a representation of the (semi)-lattice of information structures where each element is represented as a function of its corresponding value of entropy or mutual information. In abscissa are the dimensions $k$ and in ordinate the values of information functions of a given subset of k variables. \\
\begin{figure} [!h]
	\centering
	\includegraphics[height=6cm]{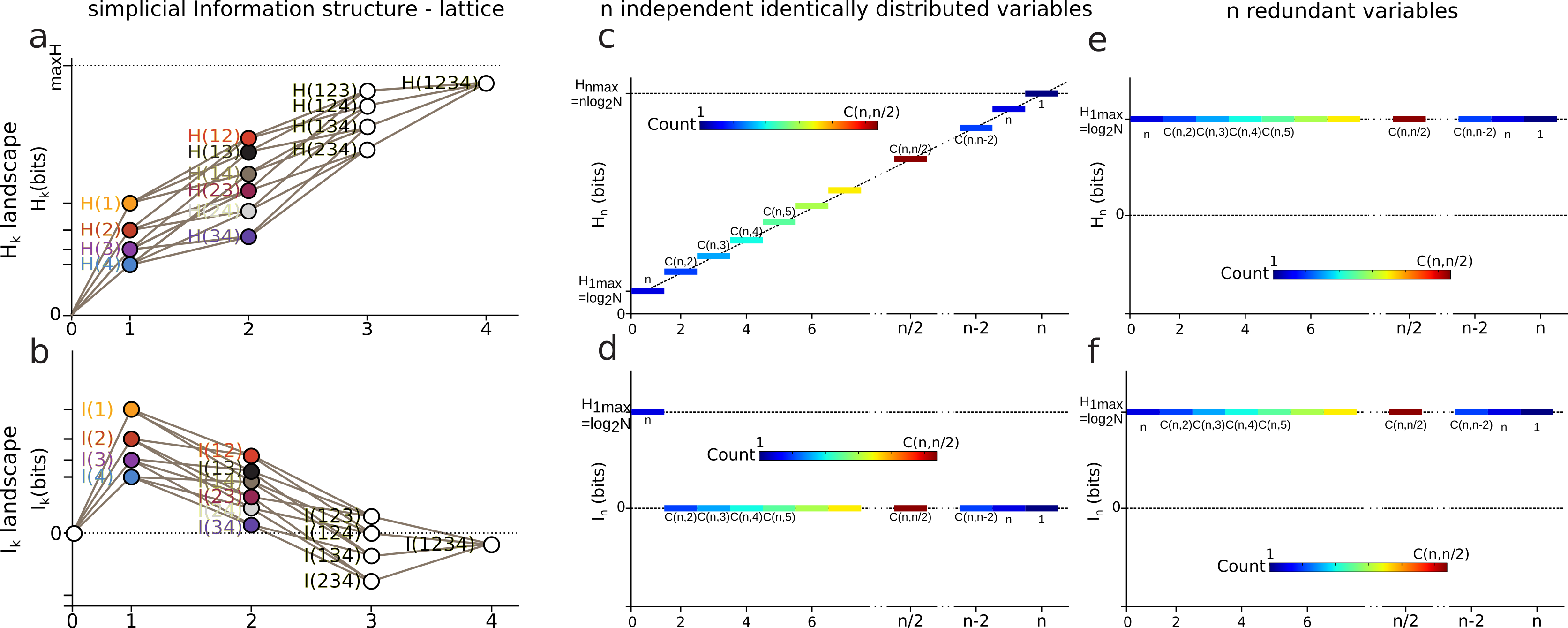}
	\caption{\textbf{Entropy and information landscapes. a,} illustration of the principle of entropy $H_k$ landscape and \textbf{b,} of a mutual-information $I_k$ landscape for $n=4$ random variables. The lattice of the simplicial information structure is depicted with grey lines.\textbf{Theoretical examples of entropy and information landscapes. c,d,}  $H_k$  and $I_k$ landscapes  for n independent identically distributed variables. The degeneracy of $H_k$  and $I_k$ values is represented by a color code: the number of k-tuples having the same information value.  \textbf{e,f}  $H_k$  and $I_k$ landscapes for n fully redundant variables. Such variables are equivalent from the information point of view, they are identically distributed and fully dependent. }
	\label{figure_Supp_infolandscape}
\end{figure}
In data science terms, these landscapes provide a visualization of the potentially high-dimensional structure of the data points. In information theory terms, it provides a representation of Shannon's work on lattice \cite{Shannon1953} further developed by Han \cite{Han1975}. $H_k$ and $I_k$, as real continuous functions, provide a ranking of the lattices at each dimension $k$. It is the ranking, i.e. the relative values of information, which matters and comes out of the homological approach, rather than the absolute values. The principle of $H_k$ and $I_k$ landscapes is illustrated in Figure \ref{figure_Supp_infolandscape} for $n=4$. $H_k$ and $I_k$ analyse quantify the variability-randomness and statistical dependences at all dimensions $k$, respectively, from $1$ to $n$, $n$ being the total number of variables under study. The $H_k$ landscape represents the values of joint entropy for all k-tuples of variables as a function of the dimensions $k$, the number of variables in the k-tuple, together with the associated edges-paths of the lattice (in grey). The $I_k$ landscape represents the values of mutual-information for all k-tuples of variables as a function of the dimension $k$, which is the number of variables in the $k$-tuple. Figure \ref{figure_Supp_infolandscape} gives two theoretical extremal examples of such landscapes, one for independent identically distributed variables (totally disordered) and one for fully dependent identically distributed variables (totally ordered). The degeneracy of  $H_k$  and $I_k$ values is given by the binomial coefficient (color code in Figure \ref{figure_Supp_infolandscape}), hence allowing to derive the normal exact expression of the information landscapes in the assymptotic infinite dimensional limit ($n\rightarrow \infty$) by application of Laplace-Lemoivre theorem. These are theoretical extremal examples:  $H_k$  and $I_k$ landscapes effectively computed and estimated on biological data with finite sample are shown in the paper \cite{Baudot2019,Baudot2018,Tapia2018}, and in practice the finite sample size ($m$) may impose some bounds on the landscapes.

\subsection{Information Paths and Minimum Free Energy Complex} 

In this section we establish that information landscapes and paths encode directly the basic equalities, inequalities and functions of information theory and  allow us to obtain the minimum free energy complex that we estimate on data.

\subsubsection{Information Paths (definition)} 

On the discrete simplicial information lattice $\Delta_k$, we define a path of degree $k$ as a sequence of edges of the lattice that begins at the least element of the lattice (the identity-constant "0"), travels along edges from vertex to vertex of increasing dimension and ends at the greatest element of the lattice of dimension $k$. Information paths are defined on both joint-entropy and meet-mutual information semi-lattices, and the usual joint-entropy and mutual-information functions are defined on each element of such paths. Entropy path and information path of degree $k$ are noted $HP_k$ and $IP_k$, respectively, and the set of all information paths is noted $\mathcal{HP}_k=\{HP_i\}_{i \in {1,...,k!}}$ for the entropy paths, and $\mathcal{IP}_k=\{IP_i\}_{i \in {1,...,k!}}$ for the mutual-information paths. We have the theorem:

\begin{thm}
	The two sets of all information paths $\mathcal{HP}_k$ and $\mathcal{IP}_k$ in the simplicial information structure $\Delta_k$ are both in bijection with the symmetric group $S_k$. Notably, there are $k!$ information paths in $\Delta_k$.
\end{thm}
Proof: by simple enumeration, an edge of dimension $m$ connects $k-m$ edges of dimension $m+1$, the number of paths is hence $(k-0).(k-1)....(k-k+2).(k-k+1)=k!$, hence the conclusion $\square$.\\
A given path can be identified with \textbf{a permutation} or \textbf{a total order} by extracting the missing variable in a previous node when increasing the dimension, for example the mutual-information path in $\Delta_4$: $IP_i=0\rightarrow (0,X_2) \rightarrow (0,X_1,X_2) \rightarrow (X_1,X_2,X_4)\rightarrow (0,X_1,X_2,X_3,X_4)$ can be noted as the permutation $\sigma$: 
\begin{equation}
\left(\begin{array}{ccccc}
0 & 1 & 2 & 3 & 4\\
0 & 2 & 1 & 4 & 3
\end{array}\right) \text{ or } (01234)\xrightarrow{\sigma}(02143)  
\end{equation}
We note an information path with arrows, giving for the previous example $IP_i=(0\rightarrow X_2\rightarrow X_1\rightarrow X_4 \rightarrow X_3)$. These paths shall be seen as the automorphisms of $\{1,2 . . . . . k\}=[k]$ and the space of entropy and mutual information paths can be endowed with the structure of two opposite symmetric groups $S_k$ and $S_k^{opp}$. The equivalence of the set of paths and symmetric group only holds for the subcase of simplicial structures, and the information paths in the lattice of partition are obviously much richer.  More precisely, the subset of simplicial information paths in the lattice of partitions corresponds to the automorphisms of the lattice. It is known that the finite symmetric group is the automorphism group of the finite partition lattice \cite{Bjorner1987}. The geometrical realization of information paths $\mathcal{IP}_k$ and $\mathcal{HP}_k$ consists in two dual permutohedron (see Postnikov \cite{Postnikov2009}), and gives the informational version of the work of Mat\'{u}\v{s} on conditional probability and permutohedron \cite{Matus2003}. 

\subsubsection{Derivatives, inequalities and conditional mutual information negativity} 

\begin{figure} [!h]
	\centering
	\includegraphics[height=4cm]{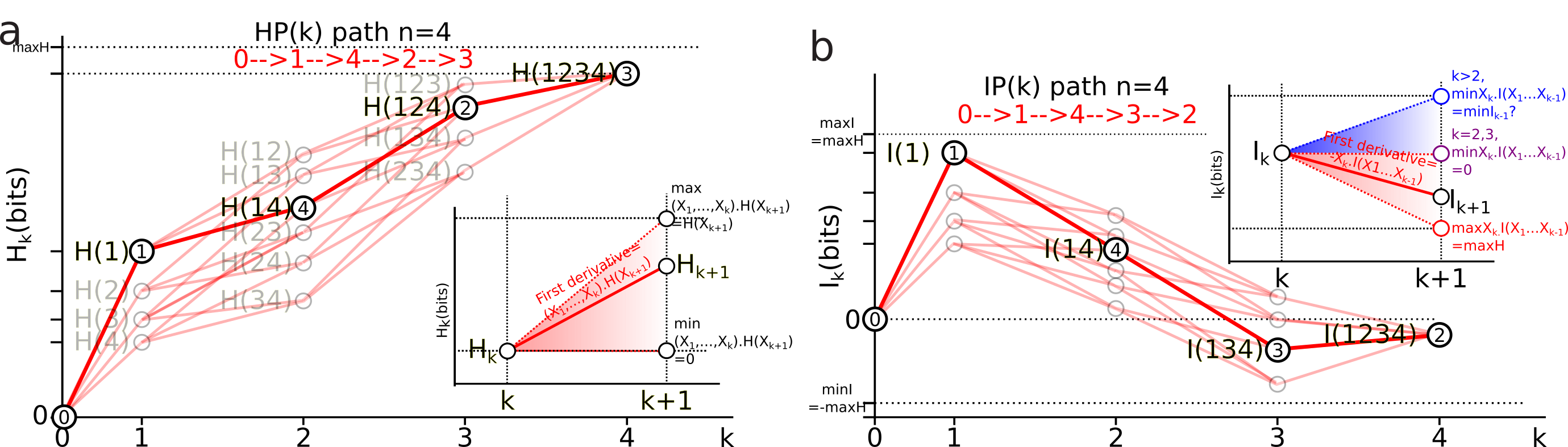}
	\caption{\textbf{Entropy and information paths.} Illustration of an entropy path $HP_i=0\rightarrow1\rightarrow4\rightarrow2\rightarrow3$ \textbf{(a)} and of a mutual information path $IP_i=0\rightarrow1\rightarrow4\rightarrow3\rightarrow2$ \textbf{(b)} for $n=4$ random variables (see text).}
	\label{figure_Supp_infopath}
\end{figure}

\paragraph{Derivatives of information paths:} 
In the information landscapes, the paths $HP_i$ and $IP_i$ are  piecewise linear functions $IP_i(k)$ with $IP_i(k)=I_k$ where $I_k$ is the mutual-information of the k-tuple of variables pertaining to the path $IP_i$.
We define the first derivatives of the paths  for both entropy and mutual information structures as piecewise linear functions:\\
\textbf{First derivative of entropy path:} the first derivative of an entropy path $HP_i(k)$ is the conditional information $(X_1,...,X_{k-1}).H(X_{k};\mathbb{P})$:
\begin{equation}\label{first derivative of entropy path}
\frac{dHP_i(k)}{dk}= H(X_1,...,X_{k};\mathbb{P}) -H(X_1,...,X_{k-1};\mathbb{P}) = (X_1,...,X_{k-1}).H(X_{k};\mathbb{P})
\end{equation}
This derivative is illustrated in the graph of Figure \ref{figure_Supp_infopath}a. It implements the chain rule of entropy  $H_{k+1}-H_{k}= (X_1;...;\widehat{X_i} ;...;X_{k+1}).H(X_i)$ (equation \ref{chain rule gener}), and in homology provides a diagram where conditional entropy is a simplicial coface map $(X_1;...;\widehat{X_i} ;...;X_{k+1}).H(X_i)=d^i:~ X^{k}\rightarrow X^{k+1}$, as a simplicial special case of Hochschild coboundaries \ref{Information structures and coboundaries}.\\
\textbf{First derivative of mutual information path:} the first derivative of an information path $IP_i(k)$ is minus the conditional information $(X_{k}).I(X_1,...,X_{k-1};\mathbb{P})$:
\begin{equation}\label{first derivative of information path}
\frac{dIP_i(k)}{dk}= I(X_1,...,X_{k};\mathbb{P}) -I(X_1,...,X_{k-1};\mathbb{P}) = -X_{k}.I(X_1,...,X_{k-1};\mathbb{P})
\end{equation}
This derivative is illustrated in the graph of Figure	\ref{figure_Supp_infopath}b. It implements the chain rule of mutual-information $I_{k-1} - I_k = X_k.I_{k-1}$ (equation \ref{chain rule gene infomut}), and in homology provides a diagram where minus the conditional mutual-information is a simplicial coface map $X_i.I(X_1;...;\widehat{X_i} ;...;X_{k+1})=d^i:~ X^{k}\rightarrow X^{k+1}$,  introduced in section \ref{Information structures and coboundaries}.

\paragraph{Bounds of the derivatives and information inequalities} 

The slope of entropy paths is bounded by the usual conditional entropy bounds  (\cite{Yeung2007} p.27-28). Its minimum is $0$ and is achieved in the case where $X_{k+1}$ is a deterministic function of ($X_1,...,X_{k}$) (lower dashed red line in Figure \ref{figure_Supp_infopath}a). Its global upperbound is $\max H_{k+1}=k.\ln (N_1...N_{k+1})$ and its sharp bound given by $(X_1;...;\widehat{X_i} ;...;X_{k+1}).H(X_i) \leq H(X_i)$ is achieved in the case where $X_{k+1}$ is independent of $X_1,...,X_{k}$ (we have $H_{k+1}=H_k+H(X_{k+1})$ (higher dashed red line in Figure \ref{figure_Supp_infopath}a). Hence, any entropy path lies in the (convex) entropy cone defined by the 3 points labeled  $H_k$, $\min H_{k+1}$ and $\max H_{k+1}$: the 3 vertices of the cone depicted as a red surface in Figure \ref{figure_Supp_infopath}a and called the Shannonian Cone following Yeung's seminal work \cite{Yeung2015}. The behavior of a mutual-information path and the bounds of its  slope are richer and more complex than the preceding conditional entropy:
\begin{itemize}  
	\item For $k=2$, the conditional information is the conditional entropy $X_i.I(X_j)=X_i.H(X_j)$ and has the same usual bounds $0\leq X_i.I(X_j)\leq I(X_j)$. 
	
	\item For $k=3$ the conditional mutual-information  $X_i.I(X_j;X_h)$ is always positive or null $X_i.I(X_j;X_h)\geq 0$\label{2-infocond posit} and hence $I_2\geq I_3$ (\cite{Yeung2007} p.26, the opposite of th.2.40 p.30), whereas the higher limit is given by $X_i.I(X_j;X_h)\geq \min\left(X_i.H(X_j),X_i.H(X_h)\right)$ (\cite{Matsuda2001} th.2.17), with equality iff $X_j$ and $X_h$ are conditionally independent given $X_i$, and  implying that the slope from $k=2$ to $k=3$ increases in the $I_k$ landscape. 
	
	\item  For $k>3$, $X_k.I(X_1;..;X_{k-1})$ can be negative as a consequence of the preceding inequalities. In terms of information landscape this negativity means that the slope is positive, hence that the information path has crossed a critical point, a minimum. As expressed by theorem \ref{infocond pos neg}, $X_k.I(X_1;..;X_{k-1})<0$ iff $I_k<I_{k+1}$. The minima correspond to zeros of conditional information (conditional independence) and hence detect cocycles in the data. The results on information inequalities define as "Shannonian" \cite{Yeung1997,Zang1997,Matus2007} the set of inequalities that are obtained from conditional information positivity ($X_i.I(X_j;X_h)\geq 0$) by linear combination, which forms a convex "positive" cone after closure. "Non-Shannonian" inequalities could also be exhibited \cite{Yeung1997}\cite{Zang1997}, hence defining a new convex cone that includes and is strictly larger than the Shannonian set. Following Yeung's nomenclature and to underline the relation with his work, we call the positive conditional mutual-information cone (surface colored in red in Figure \ref{figure_Supp_infopath}b) the "Shannonian" cone  and the negative conditional mutual-information cone (surface colored in blue  in figure \ref{figure_Supp_infopath}b) the "non-Shannonian" cone.
\end{itemize}

\subsubsection{Information paths are random processes: topological 2nd law of thermodynamic and entropy rate} 
\label{2nd law}
Here we present the dynamical aspects of information structures. Information paths provide directly the standard definition of a stochastic process and it imposes how the time arrow appears in the homological framework, how time series can be analyzed, how  entropy rates can be defined (etc.).\\
\textbf{Random (stochastic) process (definition \cite{Takacs1960}):} A random process $\{X_t,t\in T\}$ is a collection of random variables on the same probability space $(\Omega,\mathcal{B},P)$ and the index set $T$ is a totally ordered set.\\
A stochastic process is a collection of random variables indexed by time, the probabilistic version of a time series. 
Considering each symbol of a time series as a random variable, the definition of a random-stochastic process corresponds to the unique information paths $HP_i$ and $IP_i$ which total order is the time order of the series. We have the following lemma:
\begin{lemma}
	(Stochastic process and information paths): Let $(\Omega, \Delta^k, P)$ be a simplicial information structure, then the set of entropy paths  $\mathcal{HP}_k$ and of mutual-information paths  $\mathcal{IP}_k$ are in one to one correspondence with the set of stochastic processes $\{X_t,t\in T, |T|=k \}$.
\end{lemma}
Proof: direct from the definitions $\Box$.\\
As we previously stated, these paths are also automorphisms of $\{1,2 . . . . . k\}=[k]$.
We obtain immediately the  topological version of the second law of thermodynamic, which improves the result of Cover \cite{Cover1991}:\\
\begin{thm} (Stochastic process and information paths):  Let $(\Omega, \Delta^k, P)$ be a simplicial information structure, then the entropy of a stochastic process can only increase with time.
\end{thm}
Proof: given the correspondence we just established, the statement is equivalent to $H(X_1,...,X_k)\geq H(X_1,...,X_{k-1})$, which is a direct consequence of conditional entropy positivity and the chain rule of information with $k=-1/\ln2$. The generalization with respect to the stationary Markov condition used by Cover comes from the remark that in any case the indexing set of the variable is a total order.  Note that the homological formalism imposes an "initial" minimally low entropy state $H(0)=I(0)=0$ (a usual assumption in physics), the constant and zero degree homology, which has to have at least 1 component to talk about the cohomology $\Box$.\\
Remark: the meaning of this theorem in common terms was summarized by Gabor and Brillouin: \textit{"you can't have something for nothing, not even an observation"} \cite{Brillouin2014}. This increase in entropy is illustrated in Figure \label{figure_meanHk_Ik}a. The usual stochastic approach of time series assumes a Markov chain structure, imposing peculiar statistical dependences that restrict memory effects (cf. the informational characterization of Markov chains in the associated paper \cite{Baudot2019}). The consideration of stochastic processes without restriction allows any kind of dependences and arbitrary long historical and "non trivial" memory. From the biological point of view it formalizes the phenomenon of arbitrary long-lasting memory\label{memory}. From the physical point of view, without proof, such a framework appears as a classical analog of the consistent or decoherent histories developed notably by Griffiths \cite{Griffiths1984}, Omnes \cite{Omnes1988}, and Gell-Mann and Hartle \cite{Gell-Mann1990}. The information structures impose a stronger constraint of a totally ordered set (or more generally a weak ordering) than the preorder imposed by Lieb and Yngvason \cite{Lieb1998} to derive the second law. 
It is also interesting to note that even in this classical probability framework, the entropy cone (the topological cone depicted in Figure \ref{figure_Supp_infopath}a) imposed by information inequalities, when considered with this time ordering, is a time-like cone (much-like the special relativity cone), but with the arguably remarkable fact that we did not introduce any metric. \\
The stochastic process definition allows to define the finite and asymptotic information rate: 
the finite information rate $r$ of an information path $HP_i$ is $r= \frac{H_k}{k}$. The asymptotic information rate $r$ of an information path $HP_i$ is $r= \lim_{k\rightarrow\infty} \frac{H_k}{k}$. It requires the generalization of the present formalism to the infinite dimensional setting or infinite information structures, which is not trivial and will be investigated in further work. \\

\subsubsection{Local minima and critical dimension} \label{Local minima of information path} 

The derivative of information paths allows to establish the lemma on which is based the information path analysis. A critical point is said to be non-trivial if at this point the sign of the derivative of the path, i.e. the conditional information, changes.
\begin{lemma}
	\textbf{(local minima of information paths):} if $X_k.I(X_1;..;X_{k-1})<0$ then all paths from $0$ to $I_{k}$ passing by $I_{k-1}$ have at least one local minimum. In order for an information path to have a non-trivial critical point, it is necessary that $k>3$, the smallest possible dimension of a critical point being $k=3$.
\end{lemma}
Proof: it is a direct consequence of the definitions of paths and of conditional 2-mutual-information $X_k.I_2$ positivity ($X_k.I_2\geq 0$, cf. theorem \ref{2-infocond posit})$\Box$. \\

Note that by definition a local minimum can be a global minimum. We will call, if it exists, the dimension $k$ of the first local minimum of an information path the \textbf{first informational critical dimension} of the information path $IP_i$, and note it $k_{i_1}$. This allows us to define maximal information paths:\\
\textbf{Positive information path (definition):} A positive information path is an information path from $0$ to a given $I_{k}$ corresponding to a given $k$-tuple of variables such that $I_k<I_{k-1}<...<I_1$.\\
\textbf{Maximal Positive information path (definition):} A maximal positive information path is a positive information path of maximal length. More formally, a maximal positive information path  is a positive information path  that  is  not  a  proper  subset  of positive information paths. \\

The definitions make coincide positive information paths and maximal positive information path with chains (faces) and maximal chains (facets), respectively. The maximal positive information path stops at the first local minimum of an information path, if it exists.  The first informational critical dimension $k_{i_1}$ of a time series $IP_i$, whenever it exists, gives a quantification of the duration of the memory of the system.

\subsubsection{Sum over paths and mean information path} \label{Sum over paths and mean info}
As previously, for $k=1$, $IP_i(1)$ can be identified with the self-internal energy and for $k\geq2$,  $IP_i(k)$ corresponds to the k-free-energy of a single path $IP_i$. The chain rule of mutual information (equation \ref{chain rule gene infomut rec}) and the derivative of an $IP_i$ path (equation \ref{first derivative of entropy path}) implies that the k-free-energy can be obtained from a single path:
\begin{equation}
I_k=I(X_1;...;X_k;P) = I(X_1) - \sum_{i=2}^{k} X_i.I(X_1;...;X_{i-1})=IP_i(1)+ \sum_{j=2}^{k} \frac{dIP_i(j)}{dj}
\end{equation}

Hence, the global thermodynamical relation \ref{global_free} can be understood as the sum over all paths, the sum over informational histories: the classical, discrete and informational version of the path integrals in statistical physics \cite{Feynman1948}. Indeed considering an inverse relation between time and dimension $t=\frac{1}{n}$ in the probability expression \ref{Elementary_energyiid} for iid processes gives the usual expression of a unitary evolution operator  $p_{\bullet\bullet...k...\bullet}=e^{\frac{t.	E_{ij...n}}{k}}$.
Free-information-energy integrates over the simplicial structure of the whole lattice of partitions over degrees $k\geq2$, which further justifies its free energy name.\\

In order to obtain a single state function instead of a group of $k!$ paths-functions, we can compute the mean behavior of the information structure, which is achieved by defining the mean $H_k$ and $I_k$, noted $\langle H_k \rangle$ and $\langle I_k \rangle$:
\begin{equation}
\langle H_k \rangle = \frac{\sum_{T\subset [n];card(T)=k}H_k(X_T;P)}{\binom{n}{k}}
\end{equation}
and 
\begin{equation}
\langle I_k \rangle = \frac{\sum_{T\subset [n];card(T)=k}I_k(X_T;P)}{\binom{n}{k}}
\end{equation}
For example, considering $n=3$, then $\langle I_2 \rangle = \frac{I(X_1;X_2)+I(X_1;X_3)+I(X_2;X_3)}{3}$.
This defines the mean mutual information path and a mean entropy path noted  $\langle HP \rangle(k)$ and $\langle IP \rangle(k)$ in the information landscape. The case $k=2$ of those functions introduced in \cite{Baudot2018}, is studied in Merkh and Mont\'{u}far \cite{Merkh2019} with a characterization of the degeneracy of their maxima and are called Factorized Mutual-Information.
As previously, $\langle IP \rangle(1)$ can be identified with the mean self-internal energy $U(X^{n}_{hom};P_N)$ and for $k>1$  $\langle IP \rangle(k)$ to the mean k-free-information-energy $G(X^{n}_{hom};P_N)$, giving the usual isotherm relation:
\begin{equation}
H(X^{n}_{hom};P_N)=U(X^{n}_{hom};P_N)-G(X^{n}_{hom};P_N)
\end{equation}

\begin{figure} [!h]
	\centering
	\includegraphics[height=3.5cm]{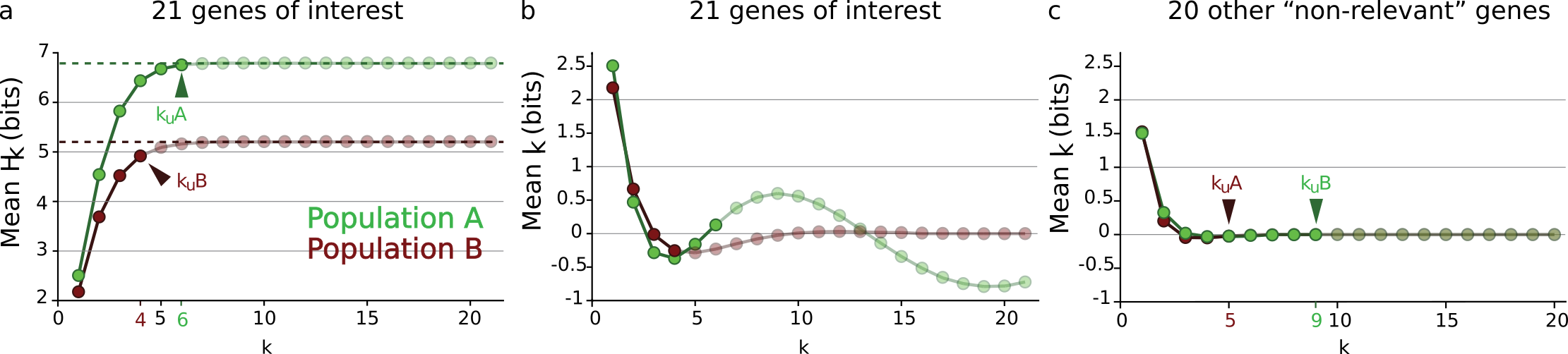}
	\caption{\textbf{Example of mean entropy and information paths of gene expression.}  \textbf{a,} Mean entropy path $\langle H_k \rangle$ for the 21 genes of interest for population A (green line) and population B neurons (red line). \textbf{b,} Mean information path $\langle I_k \rangle$ for the same pool of genes. \textbf{c,} Mean information path $\langle I_k \rangle$ for the rest of 20 genes ("non relevant"). The undersampling dimension introduced in the associated paper \cite{Baudot2019} is depicted with arrows.}
	\label{figure_meanHk_Ik}
\end{figure}
The computation of the mean paths corresponds to an idealized information structure $X^{n}_{hom}$ for which all the variables would be identically distributed, would have the same entropy, and would share the same mutual information $I_k$ at each dimension $k$: a homogeneous information structure, with homogeneous high-dimension k-body interactions. Like usually achieved in physics notably in mean-field theory (for example  Weiss \cite{Weiss1907} or Hartree), it aims to provide a single function summarizing the average behavior of the system (we will see that in practice it misses the important biological structures, pointing out the constitutive heterogeneity of biological systems see \ref{naive and non mean-field}).
Using the same dataset and results presented in \cite{Baudot2019,Baudot2018,Tapia2018}, the  $\langle IP \rangle(k)$ paths estimated on genetic expression data set are shown for two population A and population B of neurons in Figure \ref{figure_meanHk_Ik}. We quantified the gene expression levels for 41 genes in two populations of cells (A or B) as presented in \cite{Baudot2019,Baudot2018,Tapia2018}. We estimated $H_k$ and $I_k$ landscapes for these two populations and for two sets of genes ("genes of interest" and "non relevant") according to the computational and estimation methods presented in \cite{Baudot2019,Baudot2018,Tapia2018}. The available computational power restricts the analysis to a maximum of $n=21$ variables (or 21 dimensions), and imposed us to divide the genes between the two classes "genes of interest" and "non relevant". The 21 genes of interest were selected within the 41 quantified genes according to their known specific involvement in the function of population A cells. \\
Figure \ref{figure_meanHk_Ik} exhibits the critical phenomenon usually encountered in condensed matter physics, like the example of Van-der-Walls interactions \cite{Parsegian2006}. Like any $I_k$ path, $\langle IP \rangle(k)$ can have a first minimum with a critical dimension $k_{i_1}$ that could be called the homogeneous critical dimension. For the 21 genes of interest (whose expression levels, given the literature, are expected to be linked in these cell types) the  $\langle I_k \rangle$ path exhibits a clear minimum at the critical dimension $k_{i_1}=4$ for population A neurons and $k_{i_1}=5$ population B neurons, reproducing the usual free-energy potential in the condensed phase for which n-body interactions are non-negligible. For the 20 other genes, less expected to be related in these cell types, the  $\langle I_k \rangle$ path exhibits a monotonic decrease without a non-trivial minimum, which corresponds to the usual free-energy potential in the uncondensed-disordered phase for which the n-body interactions are negligible. Indeed, as shown in the work of Xie and colleagues \cite{Xie2014}, the tensor network renormalization approach of n-body interacting quantum systems gives rise to an expression of the free-energy as a function of the dimension of the interactions, in the same way than achieved here.

\subsubsection{Minimum free energy complex} \label{Minimum free energy complex} 
The analysis of information paths that we now propose aims to determine all the first critical points of information paths, in other words to determine all the information paths for which  conditional information stays positive, and all first local minima of the information landscape, that can also be interpreted as a conditional independence criterion. Such an exhaustive characterization would give a good description of the landscape and of the complexity of the measured system. The qualitative reason for considering only the first extrema for the data analysis is that, beyond that point, mutual information diverges (as section \ref{Local minima of information path} explains) and the maximal positive information paths correspond to stable functional modules in the application to data (gene expression). \\
A more mathematical justification is that they define the facets of a complex in our simplicial structure, which we will call the minimum energy complex of our information structure, underlining that this complex is the formalization of the minimum free energy principle in a degenerate case.

We now obtain the  theorem that our information path analysis aims to characterize empirically: 
\begin{thm} \label{Theorem (positive information complex)}
	\textbf{(Minimum free energy complex):} the set of all maximal positive information paths forms a simplicial complex that we call the minimum free energy complex. Moreover, the dimension-degree of the minimum free energy complex is the maximum of all the first informational critical dimensions ($d=\max k_{i_1}$), if it exists, or the dimension of the whole simplicial structure $n$. The minimum free energy complex is noted $X^{+d}$. A necessary condition for this complex not to be a simplex is that its dimension is greater or equal to four ($d\geq4$). 
\end{thm}

\begin{figure} [!h]
	\centering
	\includegraphics[height=5cm]{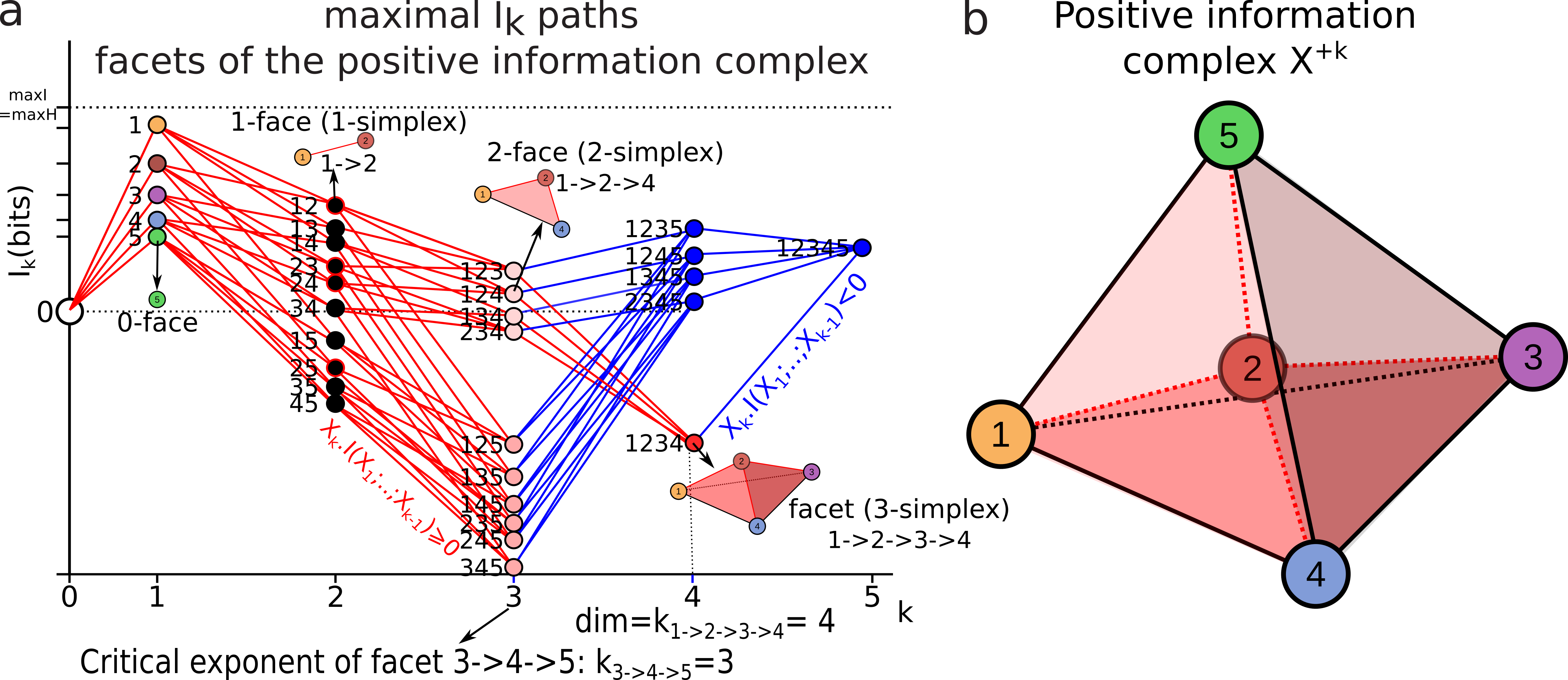}
	\caption{\textbf{Example of maximal $I_k$ paths in an $I_k$ landscape for $n=5$ together with its corresponding minimum free information energy complex. a,} maximal $I_k$ paths in an $I_k$ landscape for $n=5$. The maximum positive information paths are depicted in red, for example the paths $1\rightarrow2\rightarrow3\rightarrow4$ but also $4\rightarrow3\rightarrow2\rightarrow1$, $3\rightarrow4\rightarrow5$, and $1\rightarrow2\rightarrow5$  are maximum positive information paths, that is facets/maximal chains. 
		The facet  $1\rightarrow2\rightarrow3\rightarrow4$ is a 3-simplex while  $3\rightarrow4\rightarrow5$ is a 2-simplex with critical dimension $k_{3\rightarrow4\rightarrow5}=3$. The usual dimension of the simplex is used here, but we could have augmented it by one, since we added the constant element "0" to the algebra (pointed space), such that the usual simplicial dimension and the critical dimension correspond. The maximal critical dimension of the positive information paths is the dimension of the complex and hence $d(X^{+k})=d(1\rightarrow2\rightarrow3\rightarrow4)=4$.  \textbf{b,} The minimum free energy complex corresponding to the preceding maximal $i_k$ paths. It is a subcomplex of the 4-simplex also called the 5-cell with only one 4 dimensional cell among the five depicted as the bottom tetrahedron $\{1234\}$ with darker red volume. It has 5 vertices, 10 edges, 10 2-faces and one 3-face (cell), hence its Euler characteristic is $\chi(X^{+k})=5-10+10-1=4$ and its minimum free energy characteristic characteristic is: $H^{+k} (X^{+k})=\sum_{X_i\in X^{+k}}^5 I(X_i)-\sum_{(X_i;X_j)\in X^{+k}}^{10} I(X_i;X_j)+\sum_{(X_i;X_j;X_h)\in X^{+k}}^{10} I(X_i;X_j;X_h) - I(X_1;X_2;X_3;X4)$   }
	\label{figure_Supp_info_complex}
\end{figure}

Proof: It is known that there is a one to one correspondence between simplicial complexes and their set of maximal chains (facets) (see \cite{TaiHa2007} p.95 for example). The last part follows from Lemma 1. $\Box$.\\ 
In simple words, the maximal faces, e.g. the maximal positive information paths, encode all the structures of the minimum free energy complex. Figure \ref{figure_Supp_info_complex} illustrates one of the simplest examples of a minimum free energy complex that is not a simplex, of dimension four in a five-dimensional simplicial structure of information $\Delta_5$. \\

We define the minimum free energy characteristic as:
\begin{equation}\label{first cohomological class}
H^{+k}(X^{+k};P)=\sum_{i=1}^{k}(-1)^{i-1}\sum_{I\subset X^+;card(I)=i}I_i(X_I;P)
\end{equation} 
, where the component with dimension higher than one is a free energy. 
In the example of Figure \ref{figure_Supp_info_complex} it gives: 
\begin{equation}
\begin{split}
H^{+k}(X^{+k}) & =\sum_{X_i\in X^{+k}}^5 I(X_i)-\sum_{(X_i;X_j)\in X^{+k}}^{10} I(X_i;X_j) \\ & +\sum_{(X_i;X_j;X_h)\in X^{+k}}^{10} I(X_i;X_j;X_h) - I(X_1;X_2;X_3;X4)
\end{split}
\end{equation}
We propose that this complex defines a complex system:\\
\textbf{Complex system (definition):} \label{definition complex} A complex system is a minimum free energy complex. \\
It has the merit to provide a formal definition of complex systems as simple as the definition of an abstract simplicial complex can be, and to be quite consensual with respect to some of the approaches in this domain, as reviewed by Newman \cite{Newman2011}. Notably, it provides a formal basis to define some of the important concepts in complex systems: emergence being the coboundary map, imergence the boundary map, synergy being information negativity, organization scales being the ranks of random-variable lattices, a collective interaction being a local minimum of free-energy, diversity being the multiplicity of these minima quantified by the number of facets, a network being a 1-complex, a network of network being a 1-complex in hyper-cohomology.    \\
The interpretation in terms of sum over paths in the complex is direct as it sums over paths until conditional independence. 
We called it the minimum free energy complex but could have called it instead the positive or instantaneous complex because its facets appear as the boundaries of the "present" structure, but it obviously contains all the past-history and the memory of the structure (notably encoded in the negative $I_k$ that are necessarily non-Markovian). The topological formalization of the minimum energy allows the coexistence of numerous local minima, a situation usually encountered in complex systems (slow aging) such as frustrated glasses and K-sat problems \cite{Mezard2003,Mezard2009a} which settings correspond here to the case of $n$ binary random variables, $N_1=...=N_2=2$. The existence of the frustration effect, due to the multiplicity of these local minima in the free energy landscape \cite{Vannimenus1977}, has also been one of  the main difficulties of the condensed matter theory. Matsuda could show that $I_k$ negativity is a signature of frustration \cite{Matsuda2001}. The first axioms of DFT consider that probability densities of $n'$ elementary bodies are each in a 3-dimensional space \cite{Hohenberg1964,Kohn1965}, defining a whole simplicial structure of dimension $n=3n'$, commonly called the configuration space. When considered with the physical axiom of a configuration space, the theorem \ref{Theorem (positive information complex)} implies that, while the minimum free information energy complex of an elementary body can only be a simplex, the configuration space of $n'$ elementary bodies can be a complex with (quite) arbitrary topology. In simple terms, this settles the elementary components of the configuration space as 3-simplices, which composition can give arbitrarily complicated k-complexes. This idea is in resonance with the triangulations of space-time that arose notably from the work of Wheeler \cite{Wheeler1983a} and Penrose \cite{Penrose1971}, like spin foams \cite{Rovelli2008} and causal sets \cite{Sorkin1991}, while we only considered here classical probabilities.

\section{Discussion}
\label{Conclusion}

\subsection{Complexity through finite dimension non-extensivity and non iid} 

\paragraph{Statistical physics without statistical limit?}
The measure of entropy and information rate on data (the evolution of entropy $H_k$ when the number of variables $k$ increases) has a long history. Originally, in the work of Strong and colleagues \cite{Strong1998}, and as usual in information theory and statistical physics, it was considered that the "true" entropy was given in the asymptotic limit $\lim_{n\rightarrow \infty} H_n$ under stationarity or stronger assumptions. As explained in section \ref{biblionote on homology} (see also the note of Kontsevitch \cite{Kontsevitch1995}, in the work of Baez, Fritz and Leinster \cite{Baez2011}), and extensively in the statistical physic works of Niven  \cite{Niven2009,Niven2009a,Niven2005}, entropy does not need asymptotic or infinite assumptions such as Stirling approximation to be derived. Here  and in the associated paper \cite{Baudot2019},  we tried to understand, explore, and exploit this observation. Rather than being interested in the asymptotic limit (the infinite dimensional case) and absolute values of information, the present analysis focuses on the finite version of the "slow approach of the entropy to its extensive asymptotic limit" that Grassberger \cite{Grassberger1986}, and then Bialek, Nemenman and Tishby proposed to be "a sign of complexity" \cite{Bialek2001a}, "complexity through non-extensivity" (see also Tsallis \cite{Tsallis2002}). In short, we consider the non-extensivity of information before considering its asymptotic limit. Considering a statistical physics without statistical limit could be pertinent for the study of "small" systems, which concerns biological systems. Their small size allows them to harness thermal fluctuations, and impose their investigation with out-of equilibrium methods, as exposed in the work of Ritort and colleagues, reviewed in \cite{Ritort2008}.
The $H_k$ and $I_k$ landscapes presented here give a detailed expression of the "signs of complexity" and non-extensivity, for such small size systems (finite dimension $k$), and give a finite dimensional geometric view of the "slow approach of the entropy to its extensive asymptotic limit". In a sense, what replaces here the large number limits, Avogadro number consideration (etc.), is the combinatorial explosions of the different possible interactions: in the same way as in Van Der Walls paradigm, a combinatorial number of weak interactions can lead to a strong global interaction.  
In practice, like for any empirical investigation, the finite dimensional case imposes restrictions on the conclusions, and basically reminds us that we have not measured everything. Some relevant random variables for the observed system may be absent from the analysis, and adding such variables could reveal new interactions that are effectively constitutive and relevant to the system.
Among all possible structures of data, one is universal: data and empirical measures are discrete and finite, as emphasized by Born \cite{Born1954}, and fully justify the cohomological approach used here originating in topos (designed by Grothendieck to hold the discrete and continuous in a single hand), which was originally constructed to handle in a common framework the Lie and Galois theory, continuous and discrete symmetries.\\

\paragraph{Naive estimations let the data speaks, lack of iid or mean-field assumptions let differentiate the objects} \label{naive and non mean-field}
One of the striking results of the data analysis as presented here and in the associated paper \cite{Baudot2019} concerns the relatively low sample size ($m=41$ and $m=111$ for the analysis with cells as variable and with genes as variables respectively) required to obtain satisfying results in relatively high dimensions ($k=10$ and $k=6$ respectively). Satisfying results means here that they predict already known results reported in the biological literature, or in agreement with experts labels. In \cite{Margolin2010}, Nemenman and colleagues, who developed the problematic of the sampling problem, state in the introduction "entropy may be estimated reliably even when inferences about details of the underlying probability distribution are impossible. Thus the direct estimation of dependencies has a chance even for undersampled problems" and conclude that "a major advantage of our definition  of  statistical dependencies in terms of the MaxEnt approximations is that it can be applied even when the underlying distributions are undersampled". The present analysis agrees and confirms their conclusion. 
The method applied here is quite elementary. It  does not make assumptions of an expected or true distribution, of maximum entropy distribution or pairwise interaction Hamiltonian, coupling constant or metric, of stationarity or ergodicity or iid process, Markov chain, or underlying network structure (...) or whatever prior that would speak in place of the data. It just considers numerical empirical probabilities as expressed by Kolmogorov axioms (\cite{Kolmogorov1933a}  chap.1), which he called the "generalized fields of probability" because it does not assume the 6th axiom of continuity. Rather than fixing a model with priors, the present formalism allows the raw data to impose freely their specific structure to the model, what is usually called the naive approach or naive estimation. If one accepts that a frequentist theory and interpretation of probability is mathematically valid (\cite{Kolmogorov1933a}, chap.1), one may then conclude that a frequentist theory of entropy and information may also hold, and moreover directly fulfills the usual requirement of observability in theoretical physics recalled by Born in his Nobel lecture \cite{Born1954}. This frequentist elementary consideration is not trivial mathematically notably when considered from the number theoretic point of view. For example, the combinatoric of integer partitions of $m$ could be investigated in the general information structure (partition) context, which up to our knowledge has not been achieved in the context of probability and information. 
Up to our knowledge, all previous studies that tried to quantify statistical dependences by information methods (with more than 3 variables) used total correlation \cite{Tkacik2014,Margolin2010}, and crucially assumed that the interaction between the variables are homogeneous which corresponds to usual mean field assumption, and to the iid case of mean information \ref{Sum over paths and mean info} presented here. The combinatorial decomposition proposed allows to identify heterogeneous classes within the set of variables which would not have been possible using homogeneous assumptions. We hence believe that such combinatorial approach will play a key role in future machine learning developments and automatic classification problems. Notably, we only explored here the smallest and simplest combinatorics arising in homology, and Vigneaux already identified q-multinomial extensions of this combinatorics associated with Tsallis entropies \cite{Vigneaux2019a}. Notably, as stressed in introduction, those combinatorial decomposition can be understood as providing a geometrically constrained architecture to neural networks generalization.

\subsection{Epigenetic topological learning - biological diversity} 

\begin{figure} [!h]
	\centering
	\includegraphics[height=7.5cm]{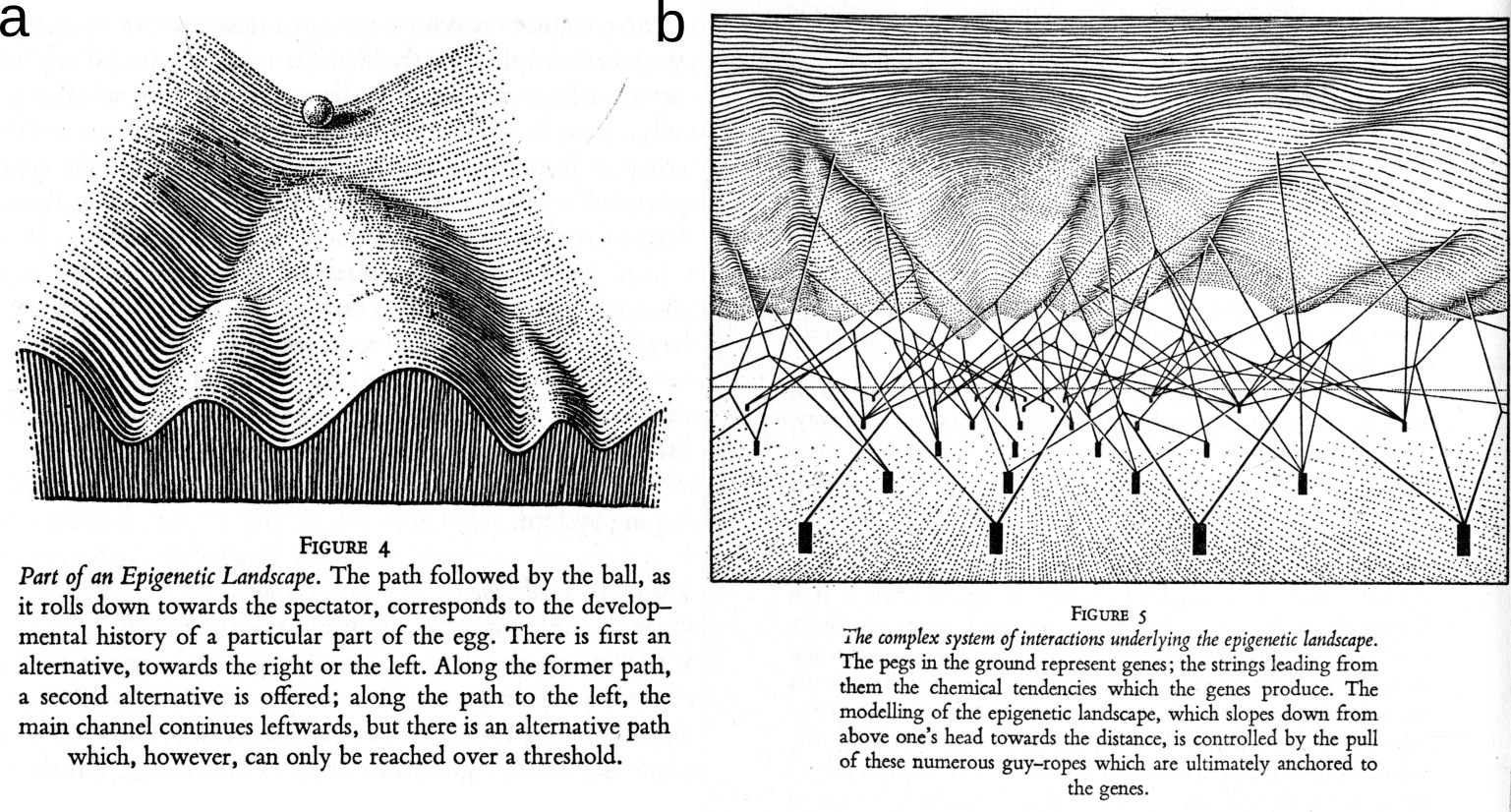}
	\caption{\textbf{The epigenetic landscape of Waddington. a,} The epigenetic landscape of Waddington, a path of the ball in this landscape illustrates a cell developmental fate. \textbf{b,} "The complex system of interactions underlying the epigenetic landscape" with Waddington's original legends \cite{Waddington1957}.}
	\label{figure_waddington}
\end{figure}

In place of the MaxEnt principle, we proposed an almost synonymous least energy principle equivalent here to a homological complex (finite and without metric assumptions). Mathematically, we took profit of the fact that whether the maximum of entropy functional is always unique and in a sense normative, the minima of $I_k$ functionals exhibit a rich structure of degeneracy, generated by the "non-Shannonian set" \cite{Yeung1997,Zang1997,Matus2007} and conjectured to be at least as rich as topological links can be. We proposed that this multiplicity of minima accounts for biological diversity, or more precisely that the number of facets of this complex quantifies the diversity in the system. The application to cell type identification presented in the associated paper \cite{Baudot2019} gives a preliminary validation of such quantification. Moreover, the definition of a complex system as the minimum free-energy complex given in section \ref{definition complex}, underlining that diversity is just the multiplicity of the minima, is in agreement with Waddington's original work \cite{Waddington1957} (see Figure \ref{figure_waddington}b). In the allegory of Waddington's epigenetic landscapes, whatever the ball, it will always fall down, a statement that can be assimilated to the second law of thermodynamic. But doing so, it will be able to take different paths: diversity comes from the multiple minima. The explanation by Waddington of such landscape is a "complex system of interactions" that can be formalized by the minimum free energy complex with interactions corresponding to the $I_k$. Moreover, formalisms assuming that the variables are identically distributed, as for the homogeneous  systems described in the section on mean paths \ref{Sum over paths and mean info}, will display a single first minima (one facet, a simplex), and hence no diversity. Sharing the same aims, Teschendorff and Enver, and then Jin and colleagues, proposed an alternative interpretation of Waddington's landscape in terms of signaling entropy \cite{Teschendorff2017} and of probability transitions \cite{Jin2018}, respectively.   

Following Thom's topological morphogenetic view of Waddington's work \cite{Thom1977}, we propose that $I_k$ landscape, paths and minimum free energy complex provide a possible informational formalization of Waddington's epigenetic complex landscape and cell fates (cf. Figure \ref{figure_waddington}). 
This formalization of Waddington's epigenetic view is consistent with the machine learning formalization of Hebbian epigenetic plasticity. From the pure formal view, the models of Hebbian neural learning like Hopfield's network, Boltzmann machines, the Infomax models proposed by Linsker, Nadal and Parga, Bell and Sejnowski \cite{Linsker1988,Nadal1999,Bell1995}) can be viewed as binary variables subcases of a generic N-ary variable epigenetic developmental process. For example, Potts model were implemented for the simulation of cell-based morphogenesis by Glazier and colleagues \cite{Chen2007}. 
Hence the topological approach can allow the treatment of neural learning and development on the ground of a common epigenetic formalism, in agreement with biological results pointing out the continuum and "entanglement" of the  biological processes underlying development and learning \cite{Galvan2010}. In terms of the current problematics of neuroscience, such generalization allows on a formal level to consider an analog coding in place of a digital coding, and the methods developed here can be applied to studies investigating (discrete) analog coding.\\
Moreover, following all the work of these last decades on the application of statistical physics to biological systems (some of them cited in this article), we propose that the epigenetic process implements the first two laws of thermodynamics, which weak topological versions are proposed to hold in the raw data space (without phase space or symplectic structure, cf. section \ref{2nd law}). As previously underlined, the condition for such an inscription of living organism dynamic into classical statistical physics to be legitimate is that the considered variables correspond to phase space variables.    \\

\section*{Previous versions}
\textbf{Previous Version:} a partial version of this work has been deposited in the method section of Bioarxiv 168740 in July 2017 and preprints \cite{Baudot2018}.\\

\section*{Acknowledgement}
This work was funded by the European Research Council (ERC consolidator grant 616827 \textit{CanaloHmics} to J.M.Goaillard) and Median Technologies, developed at Median Technologies and UNIS Inserm 1072 - Universit\'{e} Aix-Marseille, and at Institut de Mathématiques de Jussieu - Paris Rive Gauche (IMJ-PRG), and thanks previously to supports and hostings since 2007 of Max Planck Institute for Mathematic in the Sciences (MPI-MIS) and Complex System Instititute Paris-Ile-de-France (ISC-PIF). This work addresses a deep and warm acknowledgement to the researchers who  helped its realization: D.Bennequin, J.M.Goaillard, Hong Van le, G.Marrelec, M.Tapia and J.P. Vigneaux; or  supported-encouraged it: H.Atlan, F.Barbaresco, H.B\'{e}nali, P.Bourgine, F.Chavane, J.Jost, A.Mohammad-Djafari, JP.Nadal, J.Petitot, A.Sarti, J.Touboul. A partial version of this work has been deposited in the method section of Bioarxiv 168740 in July 2017 and preprints \cite{Baudot2018}.

\section*{Abbreviations}
The following abbreviations are used in this manuscript:\\
	
	\noindent 
	\begin{tabular}{@{}ll}
		iid & independent identicaly distributed\\
		$H_k$ & Multivariate k-joint Entropy\\
		$I_k$ & Multivariate k-Mutual-Information\\
		$G_k$ & Multivariate k-total-correlation or k-multi-information\\
	\end{tabular}

\bibliographystyle{acm}
\bibliography{bibtopo}

\end{document}